\documentclass[11pt,twoside,english]{article}
\usepackage[T1]{fontenc}
\usepackage[latin9]{inputenc}
\usepackage[a4paper]{geometry}
\usepackage{fancyhdr}
\usepackage{babel}
\usepackage{amsmath}
\usepackage{graphicx}
\usepackage{xcolor}
\usepackage{sidecap}
\usepackage{epstopdf}
\usepackage[unicode=true,pdfusetitle,
 bookmarks=true,bookmarksnumbered=false,bookmarksopen=false,
 breaklinks=false,pdfborder={0 0 0},pdfborderstyle={},backref=false,colorlinks=false]
 {hyperref}
\hypersetup{
 pdfpagemode=UseNone,pdfstartview=FitH,pdfborderstyle={/S/U/W 1}}
\usepackage{breakurl}

\makeatletter

\providecommand{\tabularnewline}{\\}

\usepackage[small,bf]{caption2}
\usepackage{graphicx}
\usepackage{wrapfig}
\usepackage[scaled]{helvet}
\usepackage{floatrow}

\setcaptionmargin{0cm}

\renewenvironment{thebibliography}[1]{   
\begin{oldthebibliography}{#1}     
\setlength{\itemsep}{1mm}     
\setlength{\parskip}{0em} } {   
\end{oldthebibliography} }

\makeatother

\begin{document}
\fancyhead[OL]{} 
\fancyhead[OR]{} 
\fancyhead[ER]{} 
\fancyhead[EL]{} 
\fancyfoot[EL]{}\fancyfoot[EC]{\thepage} 
\fancyfoot[OR]{}\fancyfoot[OC]{\thepage} 
\renewcommand{\headrulewidth}{0.0pt}%

\title{Moment Approximations and Model Cascades\\
for Shallow Flow}

\author{J. Kowalski\thanks{AICES Graduate School, RWTH Aachen University, email: \texttt{kowalski@aices.rwth-aachen.de}}~~and
M. Torrilhon\thanks{Department of Mathematics, RWTH Aachen University, email: \texttt{torrilhon@rwth-aachen.de}}}

\date{(2017)}
\maketitle
\begin{abstract}
Shallow flow models are used for a large number of applications including
weather forecasting, open channel hydraulics and simulation-based
natural hazard assessment. In these applications the shallowness of
the process motivates depth-averaging. While the shallow flow formulation
is advantageous in terms of computational efficiency, it also comes
at the price of losing vertical information such as the flow's velocity
profile. This gives rise to a model error, which limits the shallow
flow model's predictive power and is often not explicitly quantifiable.
 We propose the use of vertical moments to overcome this problem.
The shallow moment approximation preserves information on the vertical
flow structure while still making use of the simplifying framework
of depth-averaging. In this article, we derive a generic shallow flow
moment system of arbitrary order starting from a set of balance laws,
which has been reduced by scaling arguments. The derivation is based
on a fully vertically resolved reference model with the vertical coordinate
mapped onto the unit interval. We specify the shallow flow moment
hierarchy for kinematic and Newtonian flow conditions and present
1D numerical results for shallow moment systems up to third order.
Finally, we assess their performance with respect to both the standard
shallow flow equations as well as with respect to the vertically resolved
reference model. Our results show that depending on the parameter
regime, e.g. friction and slip, shallow moment approximations significantly
reduce the model error in shallow flow regimes and have a lot of potential
to increase the predictive power of shallow flow models, while keeping
them computationally cost efficient.
\end{abstract}


\section{Introduction}

\fancyhead[OL]{J. Kowalski and M. Torrilhon} 
\fancyhead[OR]{Moment Approximations for Shallow Flow} 
\fancyhead[ER]{J. Kowalski and M. Torrilhon} 
\fancyhead[EL]{Moment Approximations for Shallow Flow} 
\fancyfoot[EL]{}\fancyfoot[EC]{\thepage} 
\fancyfoot[OR]{}\fancyfoot[OC]{\thepage} 
\renewcommand{\headrulewidth}{0.4pt}%

Mathematical shallow flow models are successfully applied in a wide
range of scientific fields. Since early in the last century they constitute
the basis for numerical weather forecasting, an historic overview
is given in \cite{kalnay2003atmospheric}. Another traditional application
is free-surface hydraulics in rivers and channels \cite{french1985open,chanson2004environmental}.
In the last decades, new applications emerged, e.g. simulation-based
assessment of hazard due to gravity-driven mass movements such as
snow avalanches or landslides \cite{christen2010ramms,mergili2017r}.
Shallow flow models are also relevant beyond geoscience applications
and used to compute granular transport processes in chemical engineering,
or in production engineering to model coating processes (paints, printing
inks, etc), see \cite{craster2009dynamics,hagemeier2011practice}.

In all of these situations shallowness refers to the fact that the
flow's vertical extent is much smaller than its horizontal extent.
This motivates the introduction of a suitably chosen vertical average and
eventually leads to a depth-averaged model of reduced complexity.
Computational costs of shallow flow models are significantly lower
than those of a corresponding vertically resolved free-surface flow,
which is difficult to solve mainly due to the free surface. The most
famous depth-averaged flow model is certainly the shallow water system.
In its one dimensional formulation it is traditionally known as the Saint-Venant
equations. The idea itself, however, namely deriving a shallow flow
model by means of depth-averaging, is not restricted to a constitutive
equation representing water and has been applied to many other fluid
and granular flow rheologies \cite{savage1989motion,gray2014depth,baker2016two}.
Robust and efficient numerical methods have been developed to solve
them \cite{leveque2002finite,tai2001accurate}.

Quite naturally any shallow flow formulation comes at the price of
loosing vertical information, such as information on the velocity
profile. This is not critical if the velocity profile is constant
throughout the flow depth and vertical acceleration can be neglected.
In the geophysical flow context such a situation is often referred
to as 'plug-flow'. In many realistic situations, however, velocity
profiles deviate significantly from a vertically constant value. This
can be observed both in large scale field experiments \cite{kern2009measured}
as well as in small scale laboratory experiments \cite{schaefer2013velocity,sanvitale2016using}.

The computational shallow flow community encounters this obvious modeling
error by introducing the concept of a shape factor. The shape factor
is determined based on an assumed parametrization of the velocity
profile. Simple polynomial and exponential parametrizations result
in shape factors that are independent of field variables, and enter
the system as additional constants \cite{hutter2005savage,kowalski2008two}.
Though this serves as a first order correction, it is also restrictive,
as the chosen parametrization, say a linear velocity profile, is assumed
to be an appropriate choice throughout the duration of the flow. Data
acquired during transient granular chute flow experiments again indicate
that this is typically not the case and a strong regime dependency
of the velocity profile can be observed \cite{schaefer2013velocity}.
The capability to model a regime dependent velocity (and shear) profile
would be highly relevant in the geophysical context since it potentially
allows erosion rates to be calculated, as well as the coupling forces, 
both of which are of special interest from an application perspective. 

In our work we address this problem and propose a shallow flow formulation
based on vertical moments. Moment methods proved to be a powerful
tool and have been successfully applied in various scientific fields,
such as in rarefied gas dynamics \cite{Grad1949,torrilhon2016modeling}.
In the shallow flow context, moment methods allow information on the 
vertical flow structure to be preserved, while still being able to use 
depth-averaging to simplify and reduce the complexity of the system.
This idea is not entirely new and we follow the spirit of previously
published articles on secondary flow effects in curved open channel
flow and meandering rivers \cite{ghamry2005two,steffler1993depth,vasquez2006vertically},
sediment transport and mixing processes in rivers \cite{albers2007estimating,fukuoka2013toward}
as well as vertical stratification in shallow geophysical mixtures
\cite{kowalski2013shallow}. While these works focus on pragmatic
model derivations that are strongly motivated by specific applications,
the focus of our paper is a fundamental analysis and validation of
a generic model cascade shallow flows based on moment equations. 
We complement derivation and analysis of the shallow moment hierarchy
with a tailored vertically resolved reference model that will allow
us to assess the moment model's accuracy. The vertically resolved
system uses the nonlinear mapping of the shallow flow system onto
its scaled counterpart with a constant depth of unity. Such a transformation
has been used, e.g., in \cite{landau1950heat} to study thin melting
films in phase-change processes. Since then similar mappings have
been applied to a variety of processes, e.g. contact phase-change
and ablation processes \cite{lacroix1992numerical,schuller2017lagrangian}.
Comparing results of the reference system with the shallow flow moment
models allows us for the first time to analyze and discuss the model
error introduced during depth-averaging. The generic formulation of
the shallow flow moment hierarchy and its comparison to the vertically
resolved reference solutions are the major innovations of our work.

In Sec.~\ref{sec:Overview} we give a quick overview of the technical
approach of the paper. In Sec.~\ref{sec:landau}, we introduce the
coordinate mapping and derive the reference shallow flow system. It
consists of a first contribution similar to the shallow flow equation,
and a second contribution that accounts for a kinematically consistent
vertical coupling. We close this section by particularizing the reference
shallow flow system for a Newtonian constitutive relation. A moment
expansion based on the reference shallow flow system is the content
of Sec.~\ref{sec:moments}. In Sec.~\ref{sec:numex}, we present
numerical results both for the vertically resolved reference model,
and for the shallow moment systems. A different approach to treat the boundary 
conditions of the vertical velocity profile is given in Sec.~\ref{sec:strong}.
We conclude with a summary and a brief outlook. An appendix contains 
further details of the derivation and equations.

\section{Overview of the Approach\label{sec:Overview}}

In shallow flow scenarios depth-averaging typically means the loss
of information about the vertical velocity profile in $z$-direction.
The velocity components $u$ and $v$ are replaced by their integral
means $u_{m}$ and $v_{m}$. This paper tries to overcome this by
assuming a polynomial expansion of the velocity components in the
form 
\begin{align}
u(x,y,z,t) & =u_{m}(x,y,t)+\sum_{j=1}^{N}\alpha_{j}(x,y,t)\phi_{j}\left(z\right)\label{eq:generalExpansion1}\\
v(x,y,z,t) & =v_{m}(x,y,t)+\sum_{j=1}^{N}\beta_{j}(x,y,t)\phi_{j}\left(z\right)\label{eq:generalExpansion2}
\end{align}
with appropriate polynomials $\phi_{j}(z)$, $j=1,...,N$, see also
\cite{steffler1993depth}. The complete $z$-profile of $u$ and $v$
will now be modeled by the 2d functions $u_{m}(x,y,t)$ and $v_{m}(x,y,t)$
and the coefficients $\alpha_{j}(x,y,t)$ and $\beta_{j}(x,y,t)$
for $j=1,...,N$. The number $N$ will indicate the quality of the
model and we will derive concrete partial differential equations for
the coefficients for moderate values of $N$ which may replace standard
depth-averaged shallow flow models. Additionally, we give the generic
formulation of the entire model cascade. 

A crucial aspect is the validation of the models and the study of
their physical accuracy. For this purpose we derive a tailored reference
system that is able to provide synthesized data to evaluate the accuracy
of the models in numerical simulations. The reference system is based
on non-averaged Navier-Stokes equations, hence vertically resolving
the $z$-profiles of the velocity field. Both the reference system
and the reduced model cascade assumes hydrostatic equilibrium, hence
a linear pressure profile. In this way the reference system can be
used to precisely evaluate the error of vertical averaging alone.
A version of the reference system using a mapped $z$-coordinate additionally
allows a simplified derivation of the model cascade and efficient
reference computations.

\section{A Reference System for Vertically Resolved Shallow Flow\label{sec:landau}}

We use state variables velocity $\mathbf{u}=(u,v,w)^{T}$, pressure
$p$ and deviatoric stress tensor $\sigma$ as functions of space
$(x,y,z)$ and time $t$. The density $\rho$ is constant and $\mathbf{g}$
is the vector of gravitational acceleration. We allow for a general
situation, in which the $z$-axis of the coordinate system is not
necessarily collinear with the vector of gravitational acceleration.
This implies $\mathbf{g}=[g_{x},g_{y},g_{z}]^{T}=g[e_{x},e_{y},e_{z}]^{T}$,
in which $[e_{x},e_{y},e_{z}]^{T}$ is a unit vector and $g$ the
value of gravitational acceleration. The Navier-Stokes equations can
be reduced by an asymptotic analysis implied by the shallowness assumption.
Appendix \ref{sec:derivation} shows the basic arguments and, in particular,
discusses the influence of the shear stress components $\sigma_{xz}$ and $\sigma_{yz}$
of the stress tensor components, which will be important in the following. 
As a result of the asymptotic analysis, we will in this work neglect the in-plane 
deviatoric stresses $\sigma_{xx}$, $\sigma_{yy}$ and $\sigma_{xy}$. 
Note that these can still have a significant influence in certain granular flow situations, 
e.g. when they imply a cut-off frequency in the growth rate of roll waves in 
granular flow, \cite{EdwardsGray2014}, cross-slope velocity gradients due to 
wall friction \cite{baker2016two}, and regularization of segregation induced fingers 
in granular material \cite{BakerJG2016}. In these cases, the shallow moment 
cascades would have to be extended appropriately.

For now, however, we are interested in study a conceptual shallow flow moment
framework on a fundamental level, which is why the resulting system (\ref{eq:massbalcompdim})
to (\ref{eq:momzbalcompdim}) will be the starting point for the derivation
of our reference system as well as for the moment cascade. In order
to increase readability we will use the dimensional, yet reduced,
formulation of mass and momentum balances, namely
\begin{align}
\partial_{x}u+\partial_{y}v+\partial_{z}w & =0\label{eq:massbalcompred}\\
\partial_{t}u+\partial_{x}(u^{2})+\partial_{y}(uv)+\partial_{z}(uw) & =-\frac{1}{\rho}\partial_{x}p+\frac{1}{\rho}\partial_{z}\sigma_{xz}+ge_{x}\label{eq:momxbalcompred}\\
\partial_{t}v+\partial_{x}(uv)+\partial_{y}(v^{2})+\partial_{z}(vw) & =-\frac{1}{\rho}\partial_{y}p+\frac{1}{\rho}\partial_{z}\sigma_{yz}+ge_{y}\label{eq:momybalcompred}
\end{align}

\noindent The dimensional pressure is accordingly given by
\begin{align}
p(t,x,y,z)=\left(h_{b}(t,x,y)-z\right)\rho ge_{z},\label{eq:pressurered}
\end{align}
The kinematic boundary conditions (\ref{eq:kinematicbndrya}) and
(\ref{eq:kinematicbndryb}) are not affected by back-transformation
to dimensional formulation and stay unaltered.

\subsection{Mapping}

\begin{figure}
\noindent \includegraphics[width=15cm]{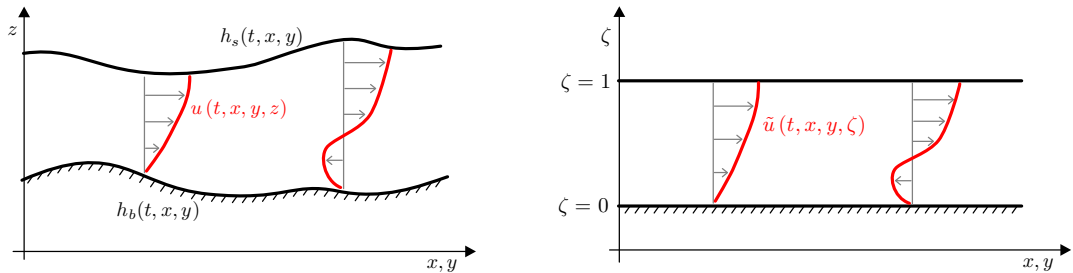}\caption{In the physical space (left plot) the flow is constrained in $z$-direction
between the basal topography at $z=h_{b}(t,x,y)$ and the free surface
at $z=h_{s}(t,x,y)$. The mapping (\ref{eq:mapping}) leads to a projected
depth coordinate $\zeta$ (right plot) and to a flow that remains
in the interval $\zeta\in[0,1]$.}
\label{fig:mapping} 
\end{figure}
The system (\ref{eq:massbalcompred})-(\ref{eq:pressurered}) already
represents our reference system, however, in order to render it more
accessible we formulate the system in terms of the scaled and normalized
vertical variable $\zeta$ defined by
\begin{align}
\zeta & =\frac{z-h_{s}(t,x,y)}{h(t,x,y)}\label{eq:mapping}
\end{align}

\noindent with the height of the flow as $h:=h_{s}-h_{b}$. This kind
of mapping immediately implies $z=h(t,x,y)\zeta+h_{b}(t,x,y)$ and
effectively transforms the free surface flow onto a domain with a
constant height of unity, see Fig.~\ref{fig:mapping}. In order to
transform the governing equations, we initially consider an arbitrary
function depending on space and time $\psi(t,x,y,z)$. Its mapped
counterpart $\tilde{\psi}(t,x,y,\zeta)$ is given by
\begin{align}
\tilde{\psi}\left(t,x,y,\zeta\right) & =\psi\left(t,x,y,h(t,x,y)\zeta+h_{b}(t,x,y)\right),
\end{align}
which implies
\begin{align}
\psi\left(t,x,y,z\right) & =\tilde{\psi}\left(t,x,y,h(t,x,y)^{-1}(z-h_{b}(t,x,y))\right).
\end{align}
Later, we want to transform the set of reduced balance laws (\ref{eq:massbalcompred})-(\ref{eq:momybalcompred}).
This requires differential operators of the mapping $\psi$, which
read
\begin{align}
h\partial_{s}\psi & =\partial_{s}\left(h\tilde{\psi}\right)-\partial_{\zeta}\left(\partial_{s}\left(\zeta h+h_{b}\right)\tilde{\psi}\right)\qquad s\in\{t,x,y\}\label{eq:transrulet}\\
h\partial_{z}\psi & =\partial_{\zeta}\tilde{\psi}\label{eq:transrulez}
\end{align}

\subsubsection{Mapping of the Mass Balance}

\noindent We multiply the mass balance (\ref{eq:massbalcompred})
with height $h$
\begin{align}
h\left(\partial_{x}u+\partial_{y}v+\partial_{z}w\right) & =0
\end{align}
and substitute for the partial derivatives according to the transformation
rules (\ref{eq:transrulet}) and (\ref{eq:transrulez}). This yields
the mapped mass balance:
\begin{align}
\partial_{x}\left(h\tilde{u}\right)+\partial_{y}\left(h\tilde{v}\right)+\partial_{\zeta}\left(\tilde{w}-\partial_{x}\left(\zeta h+h_{b}\right)\tilde{u}-\partial_{y}\left(\zeta h+h_{b}\right)\tilde{v}\right) & =0
\end{align}
The mass balance can also be written in integral form to recover an
explicit expression for the vertical velocity $\tilde{w}$:
\begin{align}
\tilde{w} & =-\partial_{x}\left(h\int_{0}^{\zeta}\tilde{u}d\hat{\zeta}\right)-\partial_{y}\left(h\int_{0}^{\zeta}\tilde{v}d\hat{\zeta}\right)+\tilde{u}\partial_{x}\left(\zeta h+h_{b}\right)+\tilde{v}\partial_{y}\left(\zeta h+h_{b}\right)\label{eq:winintegralform}
\end{align}
In order to recover the standard depth-averaged continuity equation
of the shallow water system, we subtract the vertical velocities at
the top and the bottom 
\begin{align}
\tilde{w}(t,x,y,1)-\tilde{w}(t,x,y,0)= & -\partial_{x}\left(h\int_{0}^{1}\tilde{u}d\hat{\zeta}\right)-\partial_{y}\left(h\int_{0}^{1}\tilde{u}d\hat{\zeta}\right)\nonumber \\
 & +\tilde{u}(t,x,y,1)\partial_{x}h_{s}+\tilde{u}(t,x,y,0)\partial_{x}h_{b}\\
 & +\tilde{v}(t,x,y,1)\partial_{y}h_{s}+\tilde{v}(t,x,y,0)\partial_{y}h_{b}\nonumber 
\end{align}
and substitute in the kinematic boundary conditions (\ref{eq:kinematicbndrya})
and (\ref{eq:kinematicbndryb}), which finally results in
\begin{align}
\partial_{t}h+\partial_{x}\left(hu_{m}\right)+\partial_{y}\left(hv_{m}\right) & =0.\label{eq:integartedmassbal}
\end{align}
Here, we introduced the mean velocity $u_{m}(t,x,y):=\int_{0}^{1}\tilde{u}d\zeta$,
and accordingly $v_{m}(t,x,y):=\int_{0}^{1}\tilde{v}d\zeta$, as it
is common practice for shallow flow models.

\subsubsection{Mapping of the Momentum Balance}

\noindent Similarly, we multiply the reduced momentum equation (\ref{eq:momxbalcompred})
with height $h$
\begin{align}
h\left(\partial_{t}u+\partial_{x}u^{2}+\partial_{y}(uv)+\partial_{z}(uw)\right)+h\frac{1}{\rho}\partial_{x}p & =h\frac{1}{\rho}\partial_{z}\sigma_{xz}+hge_{x}
\end{align}
and again substitute according to the transformation rules (\ref{eq:transrulet})
to (\ref{eq:transrulez}), which results in the mapped horizontal
momentum balance:
\begin{align}
\partial_{t}\left(h\tilde{u}\right)+\partial_{x}\left(h\tilde{u}^{2}\right)+\partial_{y}\left(h\tilde{u}\tilde{v}\right)\nonumber \\
+\partial_{\zeta}\left(\tilde{u}\left(\tilde{w}-\partial_{t}\left(\zeta h+h_{b}\right)-\tilde{u}\partial_{x}\left(\zeta h+h_{b}\right)-\tilde{v}\partial_{y}\left(\zeta h+h_{b}\right)\right)\right)\nonumber \\
+\frac{1}{\rho}\partial_{x}\left(h\tilde{p}\right)-\frac{1}{\rho}\partial_{\zeta}\left(\tilde{p}\partial_{x}\left(\zeta h+h_{b}\right)\right) & =\frac{1}{\rho}\partial_{\zeta}\tilde{\sigma}_{xz}+hge_{x}\label{eq:dpmomfirststep}
\end{align}
Likewise, we can map the hydrostatic pressure relation (\ref{eq:pressurered})
\begin{align}
\tilde{p}\left(x,y,t,\zeta\right) & =h\left(x,y,t\right)\left(1-\zeta\right)\rho ge_{z},
\end{align}
which allows us to simplify the pressure dependent terms in (\ref{eq:dpmomfirststep})
according to
\begin{align}
\frac{1}{\rho}\partial_{x}\left(h\tilde{p}\right)-\frac{1}{\rho}\partial_{\zeta}\left(\partial_{x}\left(\zeta h+h_{b}\right)\tilde{p}\right) & =\,\partial_{x}\left(\frac{g}{2}e_{z}h^{2}\right)+hge_{z}\partial_{x}h_{b}.\label{eq:-2}
\end{align}
Interestingly, this expression shows not explicit dependency on the
vertical coordinate any longer. Combining (\ref{eq:dpmomfirststep})
and (\ref{eq:-2}) allows to express the mapped horizontal momentum
balance concisely as
\begin{align}
\partial_{t}\left(h\tilde{u}\right)+\partial_{x}\left(h\tilde{u}^{2}+\frac{g}{2}e_{z}h^{2}\right)+\partial_{y}\left(h\tilde{u}\tilde{v}\right)+\partial_{\zeta}\left(h\tilde{u\,}\omega\left[h,\tilde{u},\tilde{v}\right]\right) & =\frac{1}{\rho}\partial_{\zeta}\tilde{\sigma}_{xz}+gh\left(e_{x}-e_{z}\,\partial_{x}h_{b}\right).\label{eq:-3}
\end{align}
The vertical coupling operator $\omega$ deserves further attention.
By substituting for the vertical velocity (\ref{eq:winintegralform}),
we find for the respective term in (\ref{eq:dpmomfirststep})
\begin{align}
 & \tilde{w}-\partial_{t}\left(\zeta h+h_{b}\right)-\tilde{u}\partial_{x}\left(\zeta h+b\right)-\tilde{v}\partial_{y}\left(\zeta h+b\right)\\
 & =-\partial_{x}\left(h\int_{0}^{\zeta}\tilde{u}d\hat{\zeta}\right)-\partial_{y}\left(h\int_{0}^{\zeta}\tilde{v}d\hat{\zeta}\right)-\zeta\partial_{t}h\nonumber \\
 & =-\partial_{x}\left(h\int_{0}^{\zeta}\tilde{u}d\hat{\zeta}\right)-\partial_{y}\left(h\int_{0}^{\zeta}\tilde{v}d\hat{\zeta}\right)+\zeta\left(\partial_{x}\left(hu_{m}\right)+\partial_{y}\left(hv_{m}\right)\right)\nonumber 
\end{align}
where $u_{m}$ and $v_{m}$ again stand for the mean velocity fields
and the integrated mass balance (\ref{eq:integartedmassbal}) has
been used. Introducing $\tilde{u}_{d}(t,x):=\tilde{u}-u_{m}(t,x)$
as well as $\tilde{v}_{d}(t,x):=\tilde{v}-v_{m}(t,x)$ for the deviation
from the true velocities $\tilde{u}$ and $\tilde{v}$ we finally
obtain
\begin{align}
h\,\omega\left[h,\tilde{u},\tilde{v}\right] & =-\partial_{x}\left(h\int_{0}^{\zeta}\tilde{u}_{d}d\hat{\zeta}\right)-\partial_{y}\left(h\int_{0}^{\zeta}\tilde{v}_{d}d\hat{\zeta}\right)\label{eq:omega}
\end{align}
Note, that we combined terms into $u_{d}$ and $v_{d}$ by writing
$\zeta=\int_{0}^{\zeta}d\hat{\zeta}$. For future reference we state
an interesting relation for $\omega$ which also clarifies its role
as vertical coupling
\begin{align}
\partial_{x}(h\tilde{u})+\partial_{y}(h\tilde{v})+\partial_{\zeta}(h\omega) & =\partial_{x}\left(hu_{m}\right)+\partial_{y}\left(hv_{m}\right)\label{eq:omegaRelation}
\end{align}
derived from (\ref{eq:omega}). 

\noindent Following the same rationale we can derive a mapped y-momentum
balance.

\subsection{Complete Reference System}

\noindent All in all, the complete vertically resolved shallow flow
system has the form
\begin{align}
\partial_{t}h+\partial_{x}\left(hu_{m}\right)+\partial_{y}\left(hv_{m}\right) & =0,\label{eq:hydbalmass}\\
\partial_{t}\left(h\tilde{u}\right)+\partial_{x}\left(h\tilde{u}^{2}+\frac{g}{2}e_{z}h^{2}\right)+\partial_{y}\left(h\tilde{u}\tilde{v}\right)+\partial_{\zeta}\left(h\tilde{u}\omega-\frac{1}{\rho}\tilde{\sigma}_{xz}\right) & =gh\left(e_{x}-e_{z}\,\partial_{x}h_{b}\right)\label{eq:hydbalmomx}\\
\partial_{t}\left(h\tilde{v}\right)+\partial_{x}\left(h\tilde{u}\tilde{v}\right)+\partial_{y}\left(h\tilde{v}^{2}+\frac{g}{2}e_{z}h^{2}\right)+\partial_{\zeta}\left(h\tilde{v}\omega-\frac{1}{\rho}\tilde{\sigma}_{yz}\right) & =gh\left(e_{y}-e_{z}\,\partial_{y}h_{b}\right),\label{eq:hydbalmomy}
\end{align}

\noindent with the vertical coupling $\omega$ that can be concisely
written as
\begin{align}
\omega & =\frac{1}{h}\overline{\partial_{x}(h\tilde{u})+\partial_{y}(h\tilde{v})}\label{eq:omega2-1}
\end{align}
based on the averaging procedure
\begin{align}
\overline{\psi} & (\zeta)=\int_{0}^{\zeta}\left(\int_{0}^{1}\psi(\check{\zeta})d\check{\zeta}-\psi(\hat{\zeta})\right)d\hat{\zeta}\label{eq:averaging-1}
\end{align}
which follows from (\ref{eq:omegaRelation}). The system is evocative
of the shallow water equations. This is true for the mass balance,
which is formulated in terms of the depth-averaged velocities $u_{m}$
and $v_{m}$. Recall however, that the momentum balance is vertically
resolved and the velocity fields $\tilde{u}$ as well as $\tilde{v}$
are functions of the vertical coordinate $\zeta$. Note also that
for a constant flow profile in $\zeta$ the vertical coupling coefficient
$\omega$ vanishes. In that case, if in addition shear stresses are
negligible $\tilde{\sigma}_{xz}=\tilde{\sigma}_{yz}=0$, the system
indeed reduces to the shallow water equations.

\subsection{Newtonian Closure for the Shallow Flow Reference System\label{sec:landaunewt}}

Our numerical examples in Sec.\,\ref{sec:numex} will realize a Newtonian
constitutive behavior, which is why we particularize the reference
formulation for this situation, see also Sec.~\ref{sec:dimanalysis}
for a discussion of the relevant contributions of the stress tensor.
For Newtonian flow the remaining basal shear components of the deviatoric
stress tensor are closed according to
\begin{align}
\sigma_{xz}=\mu\partial_{z}u\qquad\text{and}\qquad & \sigma_{yz}=\mu\partial_{z}v\label{eq:newtclosure}
\end{align}
where $\mu$ stand for the material's dynamic viscosity. Mapping the
closure relations (\ref{eq:newtclosure}) according to (\ref{eq:transrulez})
and writing them in terms of the kinematic viscosity $\nu=\mu/\rho$
yields
\begin{align}
\frac{1}{\rho}\tilde{\sigma}_{xz}=\frac{\nu}{h}\partial_{\zeta}\tilde{u}\qquad\text{and}\qquad & \frac{1}{\rho}\tilde{\sigma}_{yz}=\frac{\nu}{h}\partial_{\zeta}\tilde{v}\label{eq:ClosureNewt}
\end{align}
which closes the system (\ref{eq:hydbalmass}) to (\ref{eq:hydbalmomy})
for Newtonian flow. 

\noindent \textbf{Boundary Conditions:} In order to solve it we need
to specify dynamic boundary conditions in the form of a velocity boundary
condition both at the free-surface $h_{s}$, and at the bottom topography
$h_{b}$. At the free-surface we assume stress-free conditions
\begin{align}
\left.\partial_{z}u\right|_{z=h_{s}}=0\quad\text{and}\left.\partial_{z}v\right|_{z=h_{s}}=0\label{eq:stressfree}
\end{align}
At the basal surface, we assume slip boundary conditions in the form
\begin{align}
\left.(u-\frac{\lambda}{\mu}\sigma_{xz})\right|_{z=h_{b}}=0\quad\text{ and }\quad\left.(v-\frac{\lambda}{\mu}\sigma_{yz})\right|_{z=h_{b}}=0
\end{align}
Here, $\lambda$ stands for the slip length and $\mu^{-1}$ has been
introduced as a scaling factor for convenience, such that $\lambda$
indeed has the unit of a length scale. A vanishing slip length, $\lambda=0$,
results in a no-slip boundary condition, such as commonly applied
for Newtonian flow, whereas $\lambda\to\infty$ represents a Neumann
boundary condition with prefect slip. Intermediate values of $\lambda$
account for a mixed flow-slip behavior. Substituting for the Newtonian
closure (\ref{eq:newtclosure}) transforms the basal boundary condition
into
\begin{align}
\left.(u-\lambda\partial_{z}u)\right|_{z=h_{b}}=0\quad\text{ and }\quad\left.(v-\lambda\partial_{z}v)\right|_{z=h_{b}}=0\label{eq:stickslipmapped}
\end{align}
and after mapping according to (\ref{eq:transrulez}) the set of dynamic
boundary conditions (\ref{eq:stressfree}) and (\ref{eq:stickslipmapped})
read
\begin{align}
\left.\partial_{\zeta}\tilde{u}\right|_{\zeta=1}=0\qquad\text{and}\qquad & \left.\partial_{\zeta}\tilde{v}\right|_{\zeta=1}=0\label{eq:mappeds}\\
\left.\partial_{\zeta}\tilde{u}\right|_{\zeta=0}=\frac{h}{\lambda}\left.\tilde{u}\right|_{\zeta=0}\qquad\text{and}\qquad & \left.\partial_{\zeta}\tilde{v}\right|_{\zeta=0}=\frac{h}{\lambda}\left.\tilde{u}\right|_{\zeta=0}.\label{eq:mappedb}
\end{align}

\section{A Moment Closure for Shallow Flows\label{sec:moments}}

We use the mapped reference system for vertically resolved shallow
flow to derive a model cascade of dimensionally reduced systems based
on moment approximations.

\subsection{Averaged Momentum Balance and Polynomial Expansion}

The mapped formulation does not only render the numerical discretization
of the free-surface flow more accessible, but also allows to directly
formulate depth-averaged equations. We will consider the momentum
balance (\ref{eq:hydbalmomx}) with the Newtonian relations (\ref{eq:ClosureNewt})
but drop the tilde for readability. After integrating $\int_{0}^{1}\cdot d\zeta$
we obtain
\begin{align}\label{eq:momrawintegrated}
\partial_{t}\left(hu_{m}\right)+\partial_{x}\left(h\int_{0}^{1}u^{2}d\zeta+\frac{g}{2}e_{z}h^{2}\right)+\partial_{y}\left(h\int_{0}^{1}uvd\zeta\right)-\frac{\nu}{h}[\, \partial_\zeta u \,]^{\zeta=1}_{\zeta=0} & =hg\left(e_{x}-e_{z}\partial_{x}h_{b}\right)
\end{align}
as evolution equation for the mean velocity $u_{m}$, where we used
the fact that the vertical coupling $\omega$ vanishes at the top
and the bottom. 
By substituting for the boundary conditions (\ref{eq:mappeds}) and (\ref{eq:mappedb})
we weakly enforce the stick-slip condition
\begin{align}
\label{eq:weaklyenforcedboundary}
-\frac{\nu}{h}[\, \partial_\zeta u \,]^{\zeta=1}_{\zeta=0} = \frac{\nu}{\lambda}\left.u\right|_{\zeta=0}
\end{align}
A discussion of the use of weak boundary treatment is presented in Sec.~\ref{sec:strong} below.
Obviously, equation \eqref{eq:momrawintegrated} with \eqref{eq:weaklyenforcedboundary} is not closed as the integrals as well as the evaluation of $u$ at $\zeta=0$ can not be evaluated without
further knowledge.

\noindent We write both lateral velocity components $u$ and $v$
as a sum of the mean and its deviation, and model the deviation as
a finite polynomial expansion, that is,
\begin{align}
u(x,y,t,\zeta) & =u_{m}(x,y,t)+u_{d}(x,y,t,\zeta)=u_{m}(x,y,t)+\sum_{j=1}^{N}\alpha_{j}\left(x,y,t\right)\phi_{j}(\zeta)\label{eq:uExpand}\\
v(x,y,t,\zeta) & =v_{m}(x,y,t)+v_{d}(x,y,t,\zeta)=v_{m}(x,y,t)+\sum_{j=1}^{N}\beta_{j}\left(x,y,t\right)\phi_{j}(\zeta)\label{eq:vExpand}
\end{align}
where we use a fixed number $N$ of basis functions $\phi_{i}$ and
both the mean velocities and the coefficients $\alpha_{j}$ and $\beta_{j}$
are independent of $\zeta$. We employ scaled Legendre polynomials
for $\phi_{j}$, orthogonal on the interval $[0,1]$ and normalized
by $\phi_{j}(0)=1$. The first three polynomials are given by
\begin{align}
\phi_{1}(\zeta)=1-2\zeta,\quad\phi_{2}(\zeta)=1-6\zeta+6\zeta^{2} & ,\quad\phi_{3}(\zeta)=1-12\zeta+30\zeta^{2}-20\zeta^{3}.
\end{align}
Note, that the mean values $u_{m}$ and $v_{m}$ can be viewed as
expansion coefficient for the zeroth basis function which is constant.
Increasing the number of coefficients $N$ allows us to describe the
vertical profile of the velocity with increasing accuracy.

\noindent Using properties of the Legendre polynomials we find
\begin{align}
\int_{0}^{1}u^{2}d\zeta=u_{m}^{2}+\sum_{j=1}^{N}\tfrac{\alpha_{j}^{2}}{2j+1},\qquad & \int_{0}^{1}uvd\zeta=u_{m}v_{m}+\sum_{j=1}^{N}\tfrac{\alpha_{j}\beta_{j}}{2j+1}
\end{align}
and also the evaluation of $u$ at the bottom is easily expressed
based on (\ref{eq:uExpand}) and the normalization condition. We obtain
the equation
\begin{align}
\partial_{t}\left(hu_{m}\right)+\partial_{x}\left(h(u_{m}^{2}+\sum_{j=1}^{N}\tfrac{\alpha_{j}^{2}}{2j+1})+\frac{g}{2}e_{z}h^{2}\right) & +\partial_{y}\left(h(u_{m}v_{m}+\sum_{j=1}^{N}\tfrac{\alpha_{j}\beta_{j}}{2j+1})\right)\nonumber \\
 & =-\frac{\nu}{\lambda}(u_{m}+\sum_{j=1}^{N}\alpha_{j})+hg\left(e_{x}-e_{z}\partial_{x}h_{b}\right)
\end{align}
for $u_{m}$ and averaging the $y$-momentum balance (\ref{eq:hydbalmomy})
analogously yields$\frac{}{}$
\begin{align}
\partial_{t}\left(hv_{m}\right)+\partial_{x}\left(h(u_{m}v_{m}+\sum_{j=1}^{N}\tfrac{\alpha_{j}\beta_{j}}{2j+1})\right) & +\partial_{y}\left(h(v_{m}^{2}+\sum_{j=1}^{N}\tfrac{\beta_{j}^{2}}{2j+1})+\frac{g}{2}e_{z}h^{2}\right)\nonumber \\
 & =-\frac{\nu}{\lambda}(v_{m}+\sum_{j=1}^{N}\beta_{j})+hg\left(e_{y}-e_{z}\partial_{y}h_{b}\right)
\end{align}
for the $y$-component $v_{m}$. In case the deviations from the mean
can be neglected, all coefficients $\alpha_{i}$ and $\beta_{i}$
vanish and these equations simplify to the well-known shallow flow
equation when combined with the equation for $h$ in (\ref{eq:hydbalmass}).

\subsection{Higher Order Averages}

If the deviations can not be neglected we need expressions for the
coefficients $\alpha_{i}$ and $\beta_{i}$. By taking moments of
the velocity fields with respect to the Legendre polynomials, we obtain
\begin{align}
\int_{0}^{1}\phi_{i}ud\zeta=\tfrac{\alpha_{i}}{2i+1} & \qquad\text{and}\qquad\int_{0}^{1}\phi_{i}vd\zeta=\tfrac{\beta_{i}}{2i+1},
\end{align}
such that we can use the mapped reference equations (\ref{eq:hydbalmomx})
and (\ref{eq:hydbalmomy}) to derive moment or evolution equations
for $\alpha_{i}$ and $\beta_{i}$. The details are given in Appendix
\ref{sec:Higher-Order-Moment}. The resulting equation for $\alpha_{i}$
is given by
\begin{align}
\partial_{t}\left(h\alpha_{i}\right)+\partial_{x}\left(h(2u_{m}\alpha_{i}+\sum_{j,k=1}^{N}A_{ijk}\alpha_{j}\alpha_{k})\right)+\partial_{y}\left(h(u_{m}\beta_{i}+v_{m}\alpha_{i}+\sum_{j,k=1}^{N}A_{ijk}\alpha_{j}\beta_{k})\right)\nonumber \\
=u_{m}D_{i}-\sum_{j,k=1}^{N}B_{ijk}D_{j}\alpha_{k}-(2i+1)\frac{\nu}{\lambda}\left(u_{m}+\sum_{j=1}^{N}(1+\frac{\lambda}{h}C_{ij})\alpha_{j}\right)\label{eq:highermoments-x}
\end{align}
where the matrices $A_{ijk}$, $B_{ijk}$ and $C_{ij}$ are defined
in (\ref{eq:Aijk}), (\ref{eq:Aij(k)}) and (\ref{eq:Aij}) of the
appendix based on integrals of the Legendre polynomials. The right-hand-side
contains non-conservative terms involving the expression
\begin{align}
D_{i} & =\partial_{x}\left(h\alpha_{i}\right)+\partial_{y}\left(h\beta_{i}\right)\label{eq:divD}
\end{align}
that stems from the vertical coupling $\omega$. The analogous equation
for $\beta_{i}$ reads
\begin{align}
\partial_{t}\left(h\beta_{i}\right)+\partial_{x}\left(h(u_{m}\beta_{i}+v_{m}\alpha_{i}+\sum_{j,k=1}^{N}A_{ijk}\alpha_{j}\beta_{k})\right)+\partial_{y}\left(h(2v_{m}\beta_{i}+\sum_{j,k=1}^{N}A_{ijk}\beta_{j}\beta_{k})\right)\nonumber \\
=v_{m}D_{i}-\sum_{j,k=1}^{N}B_{ijk}D_{j}\beta_{k}-(2i+1)\frac{\nu}{\lambda}\left(v_{m}+\sum_{j=1}^{N}(1+\frac{\lambda}{h}C_{ij})\beta_{j}\right)\label{eq:highermoments-y}
\end{align}
and exhibits the same structure. Note that $N$ indicates the order of the moment model,
while $i$ denotes the actual equation for the specific moment $\alpha_i$. 
Note also, that gravitational forces like the hydrostatic pressure or the bottom topography 
do not enter these higher order moment equations.

\subsection{Examples}

In this section, we will look at three specific examples of the model
cascade, namely the moment models of zeroth, first and second order.
In order to keep things simple, we neglect any influence by the basal
topography ($h_{b}=const$) and align the vertical axis of the coordinate
system with the negative direction of gravitational acceleration ($e_{x}=e_{y}=0$
and $e_{z}=1$). Additionally, we only present one-dimensional equations.

\subsubsection{Zeroth Order System or Shallow Water Equations}

The zeroth order shallow moment system then reads
\begin{align}
\partial_{t}\left(\begin{array}{c}
h\\
hu_{m}
\end{array}\right)+\partial_{x}\left(\begin{array}{c}
hu_{m}\\
hu_{m}^{2}+g\frac{h^{2}}{2}
\end{array}\right)=-\tfrac{\nu}{\lambda}\left(\begin{array}{c}
0\\
u_{m}
\end{array}\right)\label{eq::momentsystem0}
\end{align}
For vanishing viscosity ($\nu=0$) it corresponds to homogeneous shallow
water flow, which is hyperbolic for positive heights, as the characteristic
speeds are given by $a_{1,2}=u_{m}\pm\sqrt{gh}$.

\subsubsection{First Order or Linear System}

With $s=\alpha_{1}$ indicating the linear contribution to the velocity
deviation, the first order shallow moment system reads
\begin{align}
\partial_{t}\left(\begin{array}{c}
h\\
hu_{m}\\
hs
\end{array}\right)+\partial_{x}\left(\begin{array}{c}
hu_{m}\\
hu_{m}^{2}+\frac{1}{3}hs^{2}+g\frac{h^{2}}{2}\\
2hu_{m}s
\end{array}\right)=\left(\begin{array}{c}
0\\
0\\
u_{m}\partial_{x}(hs)
\end{array}\right)-P\label{eq::momentsystem1}
\end{align}
with
\begin{align}
P & =\tfrac{\nu}{\lambda}\left(\begin{array}{c}
0\\
u_{m}+s\\
3(u_{m}+s+4\frac{\lambda}{h}s)
\end{array}\right)\label{eq::momentsystemrhs1}
\end{align}
This time the eigenvalues of the flux Jacobian are given by
\begin{equation}
\begin{aligned}a_{1,2}=u_{m}\pm\sqrt{gh+s^{2}}\qquad\text{and}\qquad a_{3}=u_{m}\end{aligned}
\label{eq:eigenspeeds1}
\end{equation}
again showing that the system is hyperbolic for positive heights.
A vanishing first moment $s=0$ leads to the pair of waves that is
also present in the shallow flow or zeroth order system.

\subsubsection{Second Order or Quadratic System}

Considering the parabolic part of the velocity profile indicted by
$\kappa=\alpha_{2}$ yields the second order shallow moment system
as
\begin{align}
\partial_{t}\left(\begin{array}{c}
h\\
hu_{m}\\
hs\\
h\kappa
\end{array}\right)+\partial_{x}\left(\begin{array}{c}
hu_{m}\\
hu_{m}^{2}+\frac{1}{3}hs^{2}+\frac{1}{5}h\kappa^{2}+g\frac{h^{2}}{2}\\
2hu_{m}s+\frac{4}{5}hs\kappa\\
2hu_{m}\kappa+\frac{2}{3}hs^{2}+\frac{2}{7}h\kappa^{2}
\end{array}\right)=
Q \,\partial_{x}\left(\begin{array}{c}
h\\
hu_{m}\\
hs\\
h\kappa
\end{array}\right) -P
\label{eq::momentsystem2}
\end{align}
with
\begin{align}
Q = \left(\begin{array}{c c c c } 0 & 0 & 0 & 0 \\ 0 & 0 & 0 & 0 \\ 
0 & 0 & u_{m}-\tfrac{\kappa}{5} & \frac{s}{5} \\
0 & 0 & s & u_{m}+\tfrac{\kappa}{7} \end{array}\right) \quad \text{and} \quad
P =\tfrac{\nu}{\lambda}\left(\begin{array}{c}
0\\
u_{m}+s+\kappa\\
3(u_{m}+s+\kappa+4\frac{\lambda}{h}s)\\
5(u_{m}+s+\kappa+12\frac{\lambda}{h}\kappa)
\end{array}\right).
\end{align}
The structure of the equations is non-unique and there is in principal some flexibility 
in how the model is split into conservative flux and additional non-conservative 
contributions of the form $Q \,\partial_x V$. We chose the current form inspired 
by the underlying physical processes, as it follows naturally from the derivation. 
The flux part in conservation form corresponds to the depth-integrated inertial 
part of the reference system, whereas the non-conservative contribution 
corresponds to the depth-integrated vertical coupling factor. 

This time, the eigenvalues of the flux Jacobian do not have a simple
closed form. They have the form $a=u_{m}+c\sqrt{gh}$ where $c$ is
any root of the polynomial
\begin{align}
c^{4}-\tfrac{10\kappa}{7}c^{3}-\left(1+\tfrac{6\kappa^{2}}{35}+\tfrac{6s^{2}}{5}\right)c^{2} & +\left(\tfrac{22\kappa^{3}}{35}-\tfrac{6\kappa s^{2}}{35}+\tfrac{10\kappa}{7}\right)c\nonumber \\
 & -\tfrac{\kappa^{4}}{35}-\tfrac{6\kappa^{2}s^{2}}{35}-\tfrac{3\kappa^{2}}{7}+\tfrac{s^{4}}{5}+\tfrac{s^{2}}{5}=0
\end{align}
in which $s$ and $\kappa$ have been scaled with $1/\sqrt{gh}$ for
better readability.

\begin{SCfigure}
\includegraphics[width=7cm]{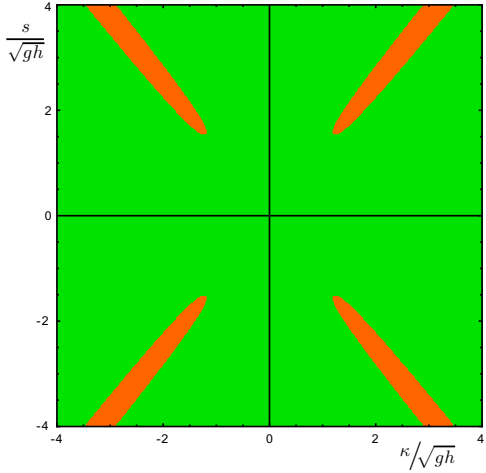}\label{fig:hyperbolicity}
\caption{Hyperbolic regions of the quadratic system parametrized in terms of the 
scaled second moment $\kappa/\sqrt{gh}$ (x-axis) and the scaled first moment 
$s/\sqrt{gh}$ (y-axis). The green area denotes the existence of a complete set of 
real valued eigenspeeds, whereas in the orange area imaginary eigenvalues occur.}
\end{SCfigure}

\noindent The hyperbolic properties of the quadratic system are shown
in Fig.~\ref{fig:hyperbolicity}. Its x-axis is given by the scaled
second moment $\kappa/\sqrt{gh}$ and the y-axis by the scaled first
moment $s/\sqrt{gh}$. Hyperbolicity, hence the existence of a mutual
and complete set of real-valued eigenspeeds, is denoted by the green
region. Orange stands for a breakdown of hyperbolicity. 

\noindent We find that the eigenvalues take a particularly
simple form for a vanishing second order moment $\kappa=0$. Then,
they reduce to
\begin{align}
a_{1,2}=u_{m}\pm\sqrt{gh+s^{2}}\qquad\text{and}\qquad a_{3,4}=u_{m}\pm\tfrac{s}{\sqrt{3}},
\end{align}
namely a fast wave pair that resembles the fast waves of the linear
system (\ref{eq:eigenspeeds1}) and an additional slow wave
pair. Also a vanishing first moment $s=0$ guarantees real eigenspeeds,
see Fig.~\ref{fig:hyperbolicity}. Note, that if both first and
second moment vanish, hence $s=\kappa=0$, we again recover the shallow
flow wave pair. 

\subsubsection{Third Order or Cubic System}
Following the same generic structure as the previous subsections allows to write the cubic system as 
\begin{align}\label{eq:cubic}
\partial_{t} V +\partial_{x} F(V) = Q\, \partial_x V - P, 
\end{align}
with $V  =\left( h, h u_{m}, hs, h \kappa, h m \right)^T$ being the vector of conserved quantities consisting of height $h$, depth-averaged velocity $u_m$, and three additional moments $s=\alpha_{1}$,$\kappa=\alpha_{2}$ as well as $m=\alpha_{3}$. Furthermore 
\begin{eqnarray}
&F(V)  =\left(\begin{array}{c} 
h u_m\\ h u^2_{m} + \tfrac{1}{3} h s^2 + \tfrac{1}{5} h \kappa^2 + \tfrac{1}{7} h m^2 + \tfrac{g}{2} h^2   \\ 
2 h u_m s + \tfrac{4}{5} h s \kappa + \tfrac{18}{35} h \kappa m \\ 
2 h u_m k + \tfrac{2}{3} h s^2 + \tfrac{2}{7} h \kappa^2 + \tfrac{4}{21} h m^2 + \tfrac{6}{7} h s m \\ 
2 h u_m m + \tfrac{6}{5} h s \kappa + \tfrac{8}{15} h \kappa m \end{array}\right), & \label{eq:cubicdetails}\\[0.3cm]
&P  =\tfrac{\nu}{\lambda}\left(\begin{array}{c}
0\\
u_{m}+s+\kappa + m \\
3\left(u_{m}+ \tfrac{h+4\lambda}{h} s +\kappa + \tfrac{h + 4 \lambda}{h} m\right)\\
5\left(u_{m}+s+ \tfrac{h + 12\lambda}{h} \kappa + m\right)\\
7\left(u_{m}+\tfrac{h + 4 \lambda}{h} s +\kappa + \tfrac{h + 24\lambda}{h} m \right)
\end{array}\right) 
\text{, }
Q = \left(\begin{array}{c c c c c} 0 & 0 & 0 & 0 & 0\\ 0 & 0 & 0 & 0 & 0\\ 
0 & 0  & u_{m}-\tfrac{\kappa}{5} & \tfrac{s}{5} - \tfrac{3 m}{35} & \tfrac{3 \kappa}{35}  \\
0 & 0  & s - \tfrac{3 m}{7} & u_{m}+\tfrac{\kappa}{7} & \tfrac{2 s}{7} + \tfrac{m}{21} \\
0 & 0  & \tfrac{6 \kappa}{5} & \tfrac{4 s}{5} + \tfrac{2 m}{15} & u_m + \tfrac{\kappa}{5} 
\end{array}\right) & \nonumber
\end{eqnarray}
Note, that the hierarchical structure of the moment model is clearly visible, when comparing $F(V)$, $Q$, and $P$ across the cascade of shallow moment systems. 

\begin{SCfigure}
\noindent \includegraphics[width=10cm]{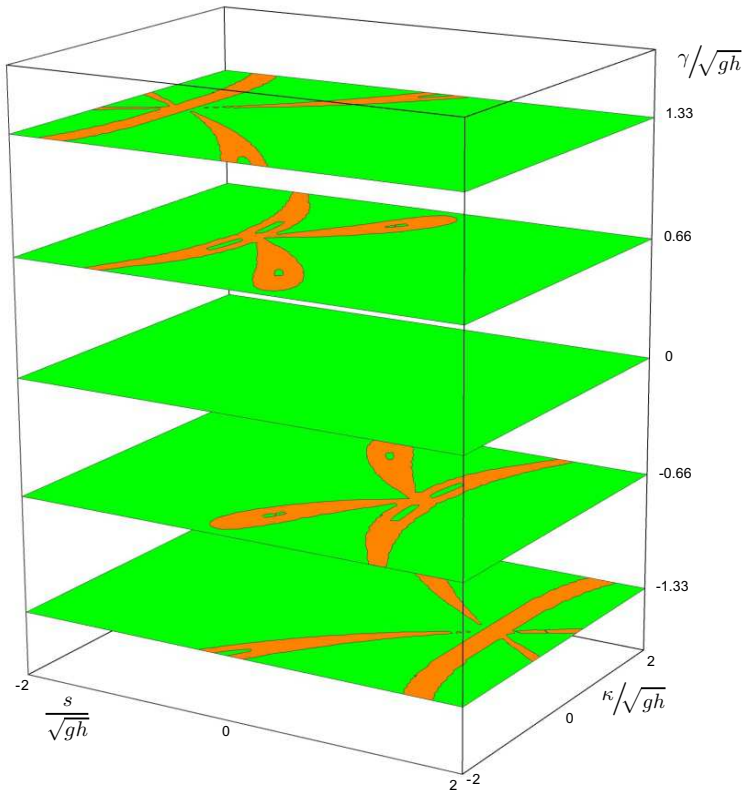}\label{fig:hyperbolicity3}
\caption{Hyperbolic regions of the cubic system parametrized in terms of the 
scaled second moment $s/\sqrt{gh}$ (x-axis), the scaled first moment 
$\kappa/\sqrt{gh}$ (y-axis) and the scaled third moment $m/\sqrt{gh}$ (z-axis). 
The green area denotes the existence of a complete set of real values eigenspeeds, 
whereas in the orange area imaginary eigenvalues occur.}
\end{SCfigure}

\noindent Similar to the quadratic system, the eigenvalues of the flux Jacobian are of
the form $a=u_{m}+c\sqrt{gh}$ where $c$ this time is the root of an order five polynomial.
The complete polynomial is given in the Appendix \ref{charpolthird}. 
The hyperbolic properties of the cubic system are shown
in Fig.~\ref{fig:hyperbolicity3}. Its x-axis is given by the scaled
second moment $\kappa/\sqrt{gh}$ and the y-axis by the scaled first
moment $s/\sqrt{gh}$ and the z-axis by the scaled thrid moment
$m/\sqrt{gh}$. Similar to the previous section, hyperbolicity is denoted 
by the green region, whereas orange stands for a breakdown of hyperbolicity. 
Again, we recover a simple form of the eigenvalues for vanishing 
higher order moments. For $\kappa=m=0$, we get
\begin{align}
a_1 = u_m, \qquad \qquad a_{2,3}=u_{m}\pm\sqrt{gh+s^{2}}\qquad\text{and}\qquad a_{4,5}=u_{m}\pm\sqrt{\frac{3}{7}}s,
\end{align}
with a fast wave pair similar to the linear
system (\ref{eq:eigenspeeds1}) that degenerates to the shallow
flow wave pair in the case of a vanishing first moment.

\subsubsection{Discussion of Hyperbolicity}

Hyperbolicity generally implies well-posedness and stability for quasi-linear, first order partial differential equations.
On the other hand, the breakdown of hyperbolicity, hence the presence of complex eigenvalues of the transport matrix in certain regions of the state space, 
can lead to uncontrolled growth of linear instabilities and unwanted grid dependencies. 
Loss of hyperbolicity is a well known challenge, e.g., for moment models 
derived from the Boltzmann equation \cite{torrilhon2016modeling} or for generalized, multi-component shallow flow models \cite{pelanti2008roe}. 
In the context of the shallow moment system we deal with in this article, the loss of hyperbolicity may be
associated with a degeneracy of the associated vertical velocity profiles or inappropriate projection of these profiles.
Note, however, that all numerical test cases presented in Section \ref{sec:numex} below have been
checked to lay well within the hyperbolic regions of their respective 
shallow moment system. Throughout the complete duration of the computation, only 
small values of the higher order moments have been observed.

\noindent When applying the shallow moment system to realistic test cases in the future,
we will have to put special emphasize on guaranteeing hyperbolicity, hence stability of the system. In particular, this is true for two-dimensional computations, in which oblique
wave vectors might lead to additional instabilities. 

The breakdown of hyperbolicity 
can be dealt with in various ways. Different projection-techniques, 
e.g. based on quadrature \cite{Fan2016}, may be the 
most promising option for a stabilizing hyperbolicity fix to the the shallow moment 
system. An alternative approach proposed in the context of kinetic gas theory
relies on a non-linear maximum entropy approach \cite{levermore1996}.

\section{Numerical Simulations\label{sec:numex}}

For the simulations presented in this section we consider $h_{b}=const$
and $e_{x}=e_{y}=0$, $e_{z}=1$ as before, and also restrict ourselves
to one-dimensional processes. The implementation of these simulations
is publicly available on GitHub \cite{ShallowFlowGithub}.

\subsection{Discretization and Setup}

The following numerical simulations will be based on equations in
dimensionless form using the scaling of Appendix \ref{sec:dimanalysis}
with Froude number $F=1$, such that the velocity scale $U$ is given
by $\sqrt{gH}$. The scaling replaces the viscosity $\nu$ by a friction
coefficient $R=\frac{\nu}{\varepsilon UH}$ representing an inverse
Reynolds number scaled by $\varepsilon$, and the slip-length $\lambda$
by the dimensionless constant $\chi=\frac{\lambda}{H}$ . The simulation
scenarios both for the reference system of vertically resolved shallow
flow equations and the moment approximations are all described by
a single setup with details given in Table \ref{table:setup}.

\begin{table}
\caption{General setup of the simulation cases for both the vertically resolved
equations and its moment approximations in Sec.~\ref{sec:numex}.}
\centering{}\global\long\def\arraystretch{2.0}
\begin{tabular}{|l|l|}
\hline
$x\in[-1,1],\quad t\in[0,t_{\text{end}}]$  & periodic domain, $t_{\text{end}}=2.0$ 
\tabularnewline
\hline
$h(x,0)=1+\exp\left(3\cos(\pi(x+x_{0}))-4\right)$  & initial shift $x_{0}=0.5$ 
\tabularnewline
\hline
$u(x,0,\zeta)=\begin{cases}
0.25 & \text{constant case }\\
0.5\zeta & \text{linear case}\\
1.5\zeta(1-\zeta) & \text{quadratic case}
\end{cases}$  & in all cases $u_{m}=0.25$ \tabularnewline
\hline
$R\in\{0,\,0.01,\,0.1\}$ (friction coefficient) & kinematic case uses $R=0$ 
\tabularnewline
\hline
$\chi\in\{0.01,\,0.1,\,\infty\}$ (slip length) & irrelevant in kinematic case 
\tabularnewline
\hline
$n_{x}=160$ (resolution), $\Delta t=0.005$ & $n_{\zeta}=80$, for reference 
system\tabularnewline
\hline
\end{tabular}\global\long\def\arraystretch{1.0}
\label{table:setup}
\end{table}

We consider a periodic domain in which the height is intialized with
a nonlinear function describing a bump that rises above the level
of $h=1$ to the value of $h\approx1.36$ in a smooth and periodic
way. The velocity $u$ in $x$-direction is constant along $x$ but
non-vanishing and we investigate three different depth profiles: constant,
linear and quadratic. The linear case corresponds to a shear flow
with zero velocity at $\zeta=0$, at least initially. The quadratic
case describes a Poiseuille-flow type profile with vanishing velocities
both at $\zeta=0$ and $\zeta=1$, again at least initially. All profiles
have a mean velocity of $u_{m}=0.25$, such that the collapse of the
height bump is super-imposed by a flow to the right. The initial conditions
are shifted to the left in order that the barycenter of the result
is centered around $x=0$ at time $t_{\text{end}}=2.0$. The values
for the friction parameter $R$ and slip length $\chi$ will be varied
in the simulation examples. We distinguish between a fully kinematic
simulation in which $R=0$ and viscous simulations with $R\neq0$.

The numerical implementation is straight forward. The moment equations
are all discretized based on the explicit high-resolution, third-order
finite-volume scheme of \cite{Cada2009,Schmidtmann2016} with explicit
time stepping using a third order stability-preserving Runge-Kutta
method. The non-conservative terms are evaluated on cell-centered
central differences together with the friction terms in an explicit
and non-split manner. The implementation has been validated on simplified
scenarios and shows robust empirical convergence upon grid refinement.

The numerical method for the reference system of vertically resolved
shallow flow equations is based on the height balance (\ref{eq:hydbalmass})
and the momentum balance (\ref{eq:hydbalmomx}) restricted to one
space dimension. Omitting the tilde for scaled variables the system
reads
\begin{align}
\partial_{t}h+\partial_{x}\left(hu\right)+\partial_{\zeta}\left(h\omega\right) & =0,\label{eq:num_h}\\
\partial_{t}\left(hu\right)+\partial_{x}\left(hu^{2}+\frac{1}{2}h^{2}\right)+\partial_{\zeta}\left(hu\omega\right) & =\frac{R}{h}\partial_{\zeta\zeta}u\label{eq:num_u}
\end{align}
where we used (\ref{eq:omegaRelation}) to replace $u_{m}$ and introduce
$\omega$ in the height balance. Per time step this system is split
into a hyperbolic left hand side solved explicitly and a diffusive
right hand side solved implicitly. For a given vertical coupling $\omega(x,\zeta,t)$
the left part of (\ref{eq:num_h})/(\ref{eq:num_u}) is discretized
as a hyperbolic system in two dimensions $(x,\zeta)$ using \cite{Cada2009,Schmidtmann2016}
again in a straight forward way. In each time step the identity $\partial_{x}\left(hu\right)+\partial_{\zeta}\left(h\omega\right)=\partial_{\zeta}\left(hu_{m}\right)$
obtained from (\ref{eq:omegaRelation}) is discretized and integrated
in $\zeta$ by trapezoidal rule on the numerical grid to obtain a
new value for the vertical coupling $\omega(x,\zeta,t)$. With this
approach also no variations in the $\zeta$-direction will be introduced
in the height field $h$. The diffusive term on the right hand side
of (\ref{eq:num_u}) is discretized implicitly by finite differences.
It uses the same cell-centered grid as the finite-volume method and
incorporates the boundary conditions
\begin{align}
\left.\partial_{\zeta}u\right|_{\zeta=0}=\frac{h}{\chi}\left.u\right|_{\zeta=0}\qquad\text{and}\qquad & \left.\partial_{\zeta}u\right|_{\zeta=1}=0\label{eq:num_bc}
\end{align}
with derivatives discretized using one-sided finite differences. As
implicit time integrator a single-diagonally implicit Runge-Kutta
method of third order is employed, see \cite{ODEbook}.

All simulations results have been obtained with a $x$-space resolution of
$160$ and constant time step $\Delta t = 0.005$. For the moment models this 
implied a CFL number between $0.5$ and $0.6$. 

\subsection{Reference Solutions\label{sec:LandauResults}}

While the actual goal of this section is the study of the moment approximations
we will first present results of the system of vertically resolved
shallow flow whose solutions will serve as reference for the moment
equations.

\subsubsection{Kinematic Case\label{sec::LandauKinematic}}

The kinematic equations result form (\ref{eq:num_h})/(\ref{eq:num_u})
by setting the friction parameter $R=0$. Boundary conditions can
then be ignored because $\omega=0$, both at the bottom and top of
the flow. It is important to note that $\omega=0$ holds for any flow
constant in $\zeta$ and the equations reduce to the classical shallow
flow equations along any line $\zeta=const$. In particular there
is no effect that generates a non-constant flow profile in $\zeta$
from a constant one in the kinematic case. Consequently, starting
with initial conditions with zero velocity the results will correspond
to classical depth-averaged shallow flow. To generate non-trivial
results for the kinematic reference equations we consider initial
conditions with non-vanishing velocities profiles but same mean velocity
as described in Table \ref{table:setup}. Recall that under such conditions
all three profiles lead to identical (averaged) initial conditions
for the standard shallow flow system, such that the impact of the
different profiles cannot be studied.

\begin{figure}
\noindent \includegraphics[width=15cm]{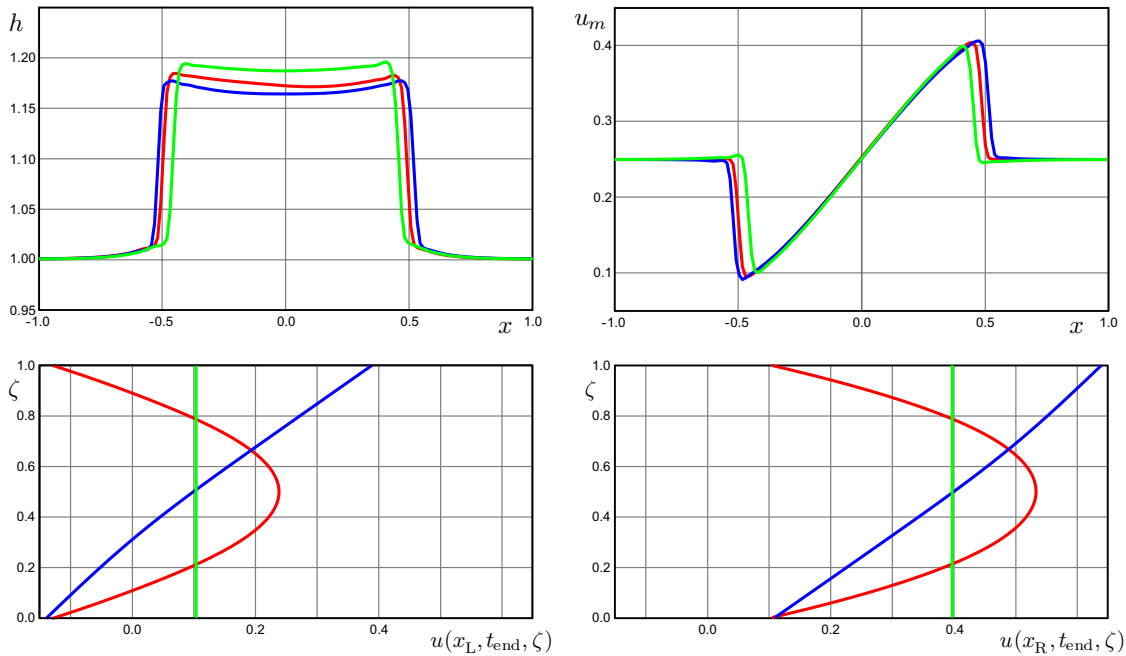}\caption{Comparison of height and mean velocity fields (top) together with
vertical velocity profiles (bottom) in purely \emph{kinematic} unsteady
solutions of the reference system for shallow flow with three different
initial velocity profiles at $t_{\text{end}}=2$, see Table \ref{table:setup}:
a \textcolor{green}{constant profile} (\textcolor{green}{green}), a \textcolor{blue}{linear profile} (\textcolor{blue}{blue}), and a \textcolor{red}{quadratic
profile} (\textcolor{red}{red}). The velocity profiles shown are taken at position $x_{L}=-0.4$
(left) and $x_{R}=0.4$ (right) of the mean velocity in the upper
right plot. }
\label{fig:LandauKinematic} 
\end{figure}

The results of the unsteady computation with the kinematic vertically
resolved equations for the three different initial velocity profiles
are shown in Fig.~\ref{fig:LandauKinematic}. The top row displays
the height and the mean velocity along the periodic domain in $x$
at $t_{\text{end}}=2.0$. The initial bump has separated into two
waves moving left and right that collide with its periodic counterparts.
At $t_{end}=2.0$ the two waves have been traveled through each other
twice and steepened into two shock waves that move towards the boundaries
of the domain. The velocity field shows the typical N-wave shape superimposed
on the constant background flow. 

The lower row shows the velocity profiles across the normalized mapped
domain $\zeta\text{\ensuremath{\in}[0,1]}$ for the three computations
at two different positions of the $x$-domain, namely $x_{L}=-0.4$
and $x_{R}=0.4$. The green curves correspond to the constant profile
case which is identical to the classical depth-averaged shallow flow
equations. It serves also as a reference to identify the influence
of the velocity profile. In the kinematic case these differences are
small but clearly visible in the results. With both the linear (blue
curves) and quadratic (red curves) initial profile the waves travel
slightly faster. Consistently, the initially constant profile yields
the maximal height at final time, while the bump for both linear and
the quadratic profile is less high and wider. In the linear case the
two shock waves remain symmetric with respect to the domain's center
as in the classical result, while the quadratic profile leads to a
slightly stronger wave on the left.

Over time the velocity profiles in $\zeta$-direction stay very similar
to the initial profile. The constant profile remains constant as discussed
above. The initially linear profiles exhibits more curvature in the
lower plots of Fig.~\ref{fig:LandauKinematic}, which indicates an
active vertical transport. The initially quadratic profile remains
vertically symmetric, however, the amplitude of the parabola around
its mean varies between the two profile plots in the figure.

\subsubsection{Friction Case\label{sec::LandauFrictionCase}}

\begin{figure}
\noindent \includegraphics[width=15cm]{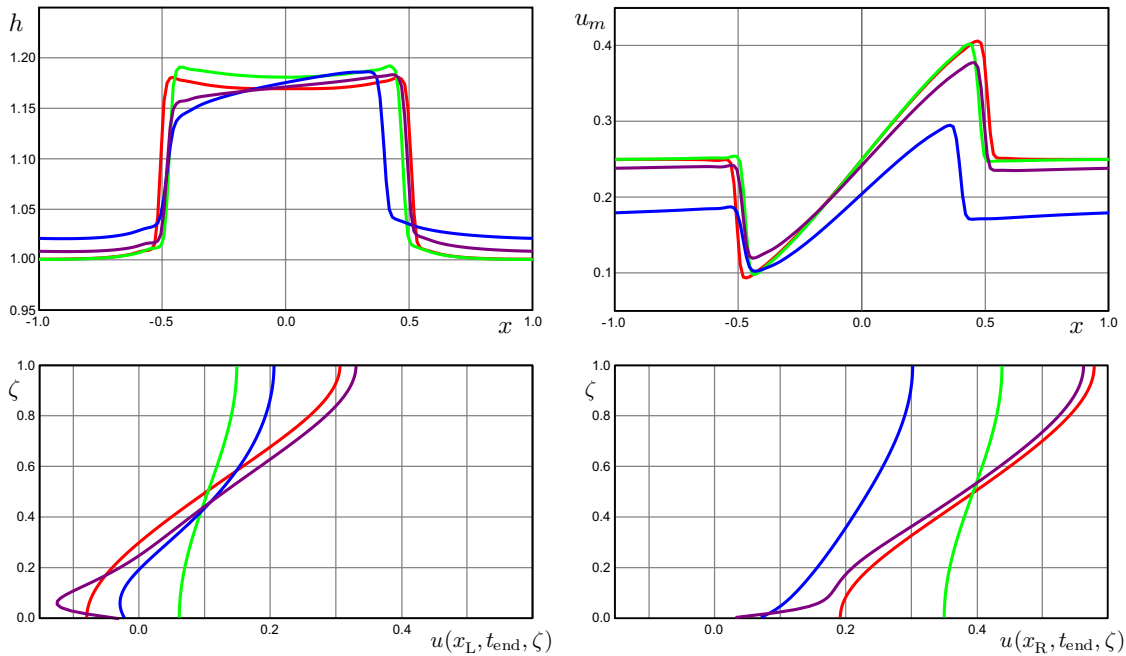}\caption{Comparison of height and mean velocity fields (top) together with
vertical velocity profiles (bottom) in \emph{viscous} unsteady solutions
of the reference system for shallow flow with \emph{linear} initial
velocity profile at $t_{\text{end}}=2$, see Table \ref{table:setup},
and four different setups of friction: force-free bottom, i.e., $\chi=\infty$,
with $R=0.01$ (\textcolor{red}{red}) and $R=0.1$ (\textcolor{green}{green}), finite slip length with
the cases $R=\chi=0.1$ (\textcolor{blue}{blue}) and $R=\chi=0.01$ (\textcolor{violet}{purple}). The velocity
profiles shown are taken at position $x_{L}=-0.4$ (left) and $x_{R}=0.4$
(right) of the mean velocity in the upper right plot. }
\label{fig:LandauFriction} 
\end{figure}

To study the effect of friction we consider the same setup as before,
see Table \ref{table:setup}, but restricted to linear initial profiles.
The results are shown in Fig.~\ref{fig:LandauFriction} with the
plots arranged in the same way as in Fig.~\ref{fig:LandauKinematic}
above. The curves only differ in the values of friction and slip coefficients
$R$ and $\chi$.

It is instructive to start with the case of a force-free bottom, which
is formally obtained by $\chi\to\infty$, such that we have Neumann
boundary conditions on either side of $\zeta$, see (\ref{eq:num_bc}).
Given by a non-vanishing parameter $R$, dissipation now leads to
an internal equilibration of the initially linear velocity profile
that drives the system towards a constant profile case. The red and
green curves in Fig.~\ref{fig:LandauFriction} correspond to force-free
bottom simulations with friction coefficients $R=0.01$ and $R=0.1$.

Due to friction the profiles become s-shaped during the simulation.
The stronger friction case (green) shows an almost constant profile
at the end of the simulation (lower row) and the height and mean velocity
fields are very similar to the classical shallow flow simulation in
Fig.~\ref{fig:LandauKinematic}. On the other hand, in the lower
friction case the linear profile is still recognizable and the result
for height and mean velocity shows almost no differences to the friction-less
case of Fig.~\ref{fig:LandauKinematic}.

Introducing the slip-flow boundary conditions (\ref{eq:num_bc}) with
$\chi\neq0$ significantly influences both the velocity profiles and,
consequently, the height and mean velocity fields. The blue and purple
curve of Fig.~\ref{fig:LandauFriction} show two results with slip
conditions. To maximize the variety of the resulting curves we choose
$R=\chi=0.1$ (blue curve) and $R=\chi=0.01$ (purple curve). Especially
the result with lower slip coefficient shows a distinct boundary layer
in which the velocity profile is pushed towards the value $u=0$.
Consequently, the velocity profile exhibits a strongly nonlinear behavior
that clearly influences also the curves of height and mean velocity.
In tendency, the slip-flow boundary leads to shock waves traveling
asymmetrically to the left and right. The force at the bottom also
leads to an energy decay and the velocity is relaxing to zero with
time.

\subsection{Moment Approximations}

\begin{figure}
\noindent \includegraphics[width=15cm]{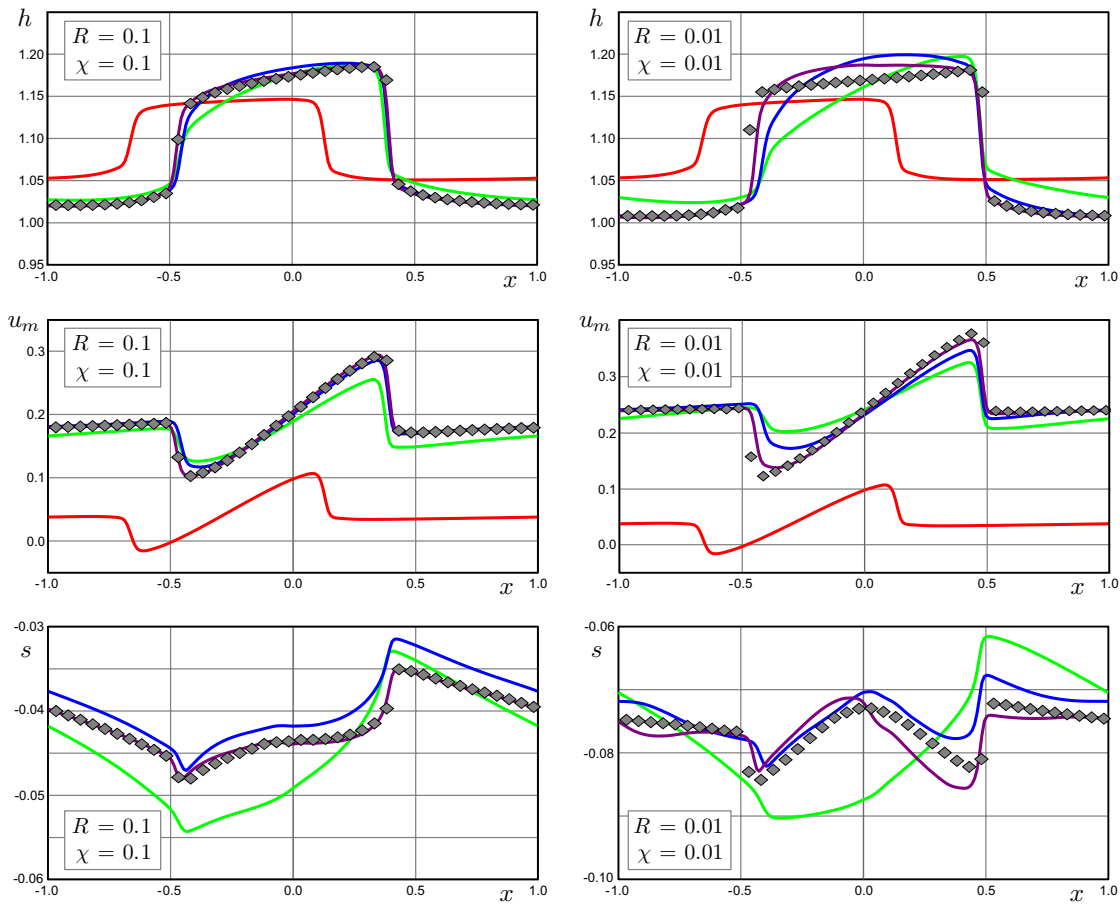}\caption{Moment approximations for shallow flow in \emph{viscous} unsteady
solutions with linear initial velocity profiles at $t_{\text{end}}=2$,
see Table \ref{table:setup}. The left column uses $R=\chi=0.1$,
and the right column uses $R=\chi=0.01$. The rows show height, mean
velocity and the first moment. Symbols represent the result of the
vertically resolved reference simulation, while the lines represent
\textcolor{red}{classical shallow flow equations} (\textcolor{red}{red}), as well as the \textcolor{green}{linear} (\textcolor{green}{green}),
the \textcolor{blue}{quadratic} (\textcolor{blue}{blue}) and the \textcolor{violet}{cubic} (\textcolor{violet}{purple}) moment system. }
\label{fig:MomentsFriction} 
\end{figure}

In order to study the convergence of the model cascade for shallow
flow we consider the same general setup as before with an initial
bump that separates and interacts with its periodic counterparts,
see Table \ref{table:setup}. The results will be displayed in Fig.~\ref{fig:MomentsKinematic}
and Fig.~\ref{fig:MomentsFriction} that follow the same plot arrangement
as above. The test cases are solved by shallow moment systems up to
third order: zeroth order system (red curves, Eqn.\,\ref{eq::momentsystem0}),
linear system (green curves, Eqn.~\ref{eq::momentsystem1}), quadratic
order (blue curves, Eqn.\,\ref{eq::momentsystem2}), and cubic order
system (purple curves). The symbols in the plots denote the reference
solution sampled from results of the vertically resolved reference
solutions discussed in Sec.~\ref{sec:LandauResults}. Higher order
moments are reconstructed from the vertically resolved velocity profile.

\noindent Note, that the zeroth order system corresponds to the classical
shallow flow system. For each test case we can hence assess the moment
hierarchy's performance with respect to both the standard shallow
flow model (red) as well as with respect to the vertically resolved
reference solution (symbols). As a rough estimate of the computational efficiency of the moment approximations we state the following relative run times for the test
cases of Table 1:
\noindent \begin{center}
\vspace{-2mm}
\begin{tabular}{|l||c|c|c|c||c|}
\hline
Theory & Shallow Flow & Linear System & Quadratic & Cubic & Reference 
System\tabularnewline
\hline
Run Time & $1.0$ & $1.7$ & $2.3$ & $3.8$ & $\approx 100.0$\tabularnewline
\hline
\end{tabular}\vspace{-2mm}
\par\end{center}
\noindent Note, that both moment, and reference implementation may
allow further performance optimizations.

\subsubsection{Kinematic Case\label{sec::MomentKinematic}}

\begin{figure}
\noindent \includegraphics[width=15cm]{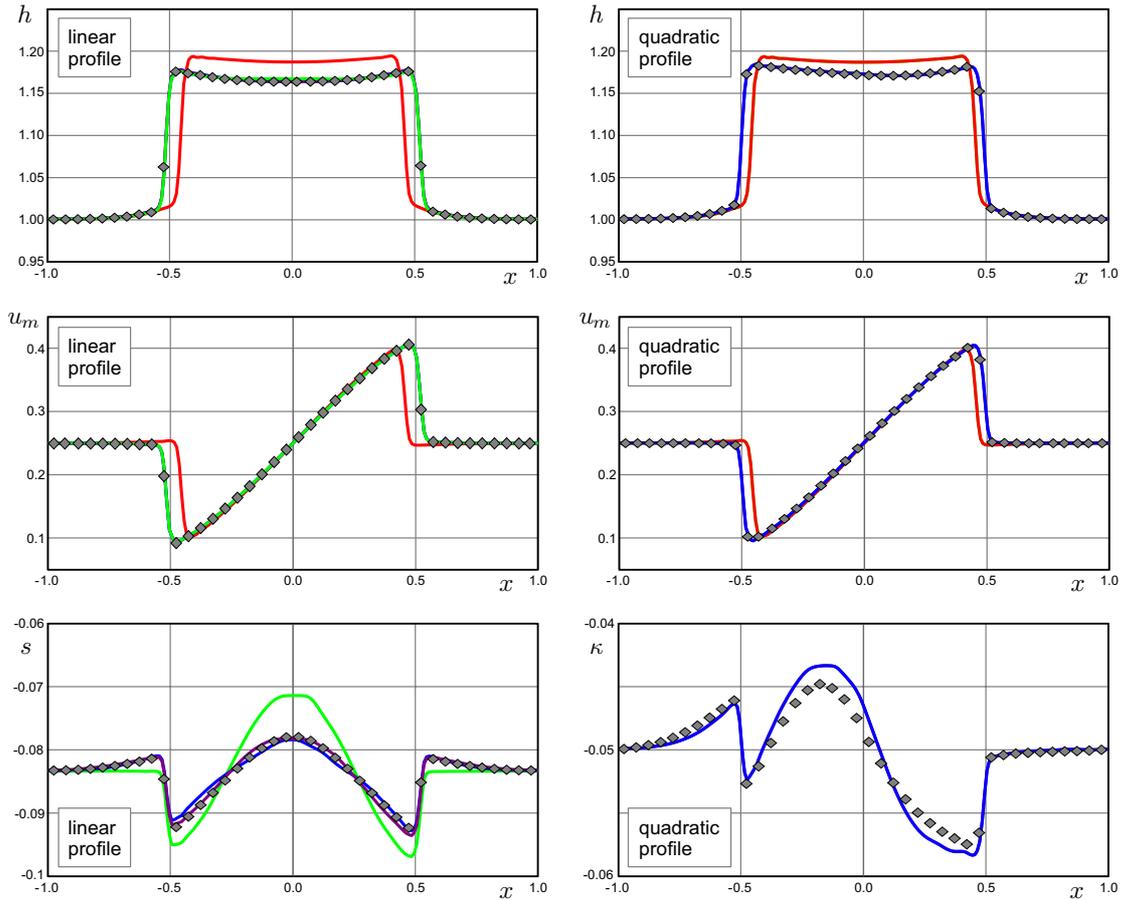}\caption{Moment approximations for shallow flow in purely \emph{kinematic}
unsteady solutions with linear (left column) and quadratic (right
column) initial velocity profiles at $t_{\text{end}}=2$, see Table
\ref{table:setup}. The height and mean velocity are shown in the
first and second row, and the bottom row displays the fields of first
and second moment, respectively. Symbols represent the result of the
vertically resolved reference simulation, while the lines represent
\textcolor{red}{classical shallow flow equations} (\textcolor{red}{red}), as well as the \textcolor{green}{linear} (\textcolor{green}{green}),
the \textcolor{blue}{quadratic} (\textcolor{blue}{blue}) and the \textcolor{violet}{cubic} (\textcolor{violet}{purple}) moment system.}
\label{fig:MomentsKinematic} 
\end{figure}

Resembling the structure of the previous discussion in Sec.~\ref{sec:LandauResults},
we consider the kinematic case ($R=0$) first. Numerical results are
shown in Fig.~\ref{fig:MomentsKinematic}. The left column displays
the results for a linear initial velocity profile, whereas the right
column shows the results of a parabolic initial velocity profile.
Top and second row show height and mean velocity fields. For the linear
initial velocity profile (left column) we additionally display the
first moment in the bottom row. As discussed earlier, see Sec.~\ref{sec::LandauKinematic},
a parabolic initial velocity profile does not generate any linear
contribution in the kinematic case such that the first moment remains
identically zero. Consequently, it is instructive to analyze the second
moment for an initially parabolic velocity profile (bottom right).

While for the linear initial velocity profile the zeroth order moment
system (shallow water system) differs clearly visible from the reference
solution, we observe that linear and higher order systems are almost
perfectly aligned with the reference solution for height and mean
velocity. Only the height (upper left) of the linear system (green)
deviates slightly around the domain's center. Numerical results of
the first moment (bottom left) allow further insight. Note, that the
first moment of the zeroth order system naturally vanishes such that
there is no red curve in this plot. The remaining curves nicely show
convergence of the moment hierarchy with increasing order. While the
linear system still over- and undershoots the reference solution significantly,
the quadratic and especially the cubic order system capture the first
moment pattern much better.

For the initially parabolic velocity profile we again observe that
height and mean velocity of the zeroth order system differ significantly
from the reference solution. Only this time, we cannot see any improvement
by using the linear system. In fact both overlap exactly, that is,
the green curves lies behind the red. This makes sense as for a zero
first moment $s=0$ the linear system (\ref{eq::momentsystem1}) reduces
to the zeroth order system (\ref{eq::momentsystem0}). The quadratic
system agrees very nicely with the reference solution, while the cubic
result is again identical to the quadratic. This time the second moment
(lower right) allows further insight. It naturally vanishes in both
zeroth and first order system (absence of red and green curve in the
plot). The second moment calculated with the quadratic order system
captures the reference solution nicely except for occasional over-
and undershooting. The cubic order system yields no further improvement,
and simply reproduces the results of the quadratic system such that
the purple curve is covered in the plot. In this case further improvement
is only expected for the fourth order moment system, which is not
considered in our work.

\subsubsection{Friction Case}

The full benefit of the shallow moment hierarchy becomes evident when
including friction. Here, we concentrate on two slip-flow parameter
settings corresponding to $R=\chi=0.1$ (left column) and $R=\chi=0.01$
(right column). In both test cases the initial velocity profile is
linear. The corresponding vertically resolved reference solution has
been discussed in Sec.~\ref{sec::LandauFrictionCase}. The results
of the moment models is shown in Fig.~\ref{fig:MomentsFriction}.
In each column the plots are given by height, mean velocity and first
moment in the top, middle and bottom rows. We observe that the zeroth
order moment system (classical shallow flow theory in red) cannot
capture the bump's shape at final time adequately at all. For neither
of the two parameter settings it can reproduce the flow's height,
its mean velocity, or the position of the bump at final time.

\noindent The larger slip length case $\chi=0.1$ (left column) already
benefits significantly from using the linear system (green). Height
and position of the right-going shock wave match the reference solution.
The plot of the first moment (bottom left), however, reveals that
there is still a significant discrepancy around the left going wave.
This error gets smaller when increasing the order of the moment system
(blue and purple curves). As expected it is the cubic system that
yields the most accurate result, indicating a converging behavior
of the moment hierarchy.

\noindent As we decrease the slip length $\chi=0.01$ (right column)
the linear system (green) can still correctly capture the position
of the right-going wave, but it predicts the mean velocity field less
accurately, especially maximum and minimum velocities. Considering
the first moment (bottom right) the situation gets even worse. Here,
the linear system even fails to qualitatively capture the behavior.
The linear system results in an N-shaped profile, while the reference
solution shows an W-shaped first moment that indicates the presence
of strong nonlinearities given by the boundary layer, see Sec.~\ref{sec::LandauFrictionCase}.
Again, we observe that the quadratic, and eventually the cubic system
increase model accuracy and correctly account for the aforementioned
W-shape of the first moment. The absolute model error is however larger
with decreasing slip length.

\section{Strong Enforcement of Boundary Conditions}\label{sec:strong}

The boundary conditions of the velocity profile at the bottom and
at the free surface have only been weakly enforced in the moment approximations
of Sec.~\ref{sec:moments} using partial integration of the friction
terms. It is also possible to impose the boundary conditions (\ref{eq:mappeds})/(\ref{eq:mappedb})
directly onto the polynomial ansatz (\ref{eq:uExpand})/(\ref{eq:vExpand}).
This gives the linear equations 
\begin{align}
\sum_{j=1}^{N}j(j+1)(-1)^{j}\alpha_{j}=0,\qquad & \sum_{j=1}^{N}(j(j+1)\lambda+h)\alpha_{j}=-hu_{m}\label{eq:constraints}
\end{align}
as constraints to the variables $\{\alpha_{j}\}_{j=1,...N}$. An analogous
expression holds for the variables $\{\beta_{j}\}_{j=1,...N}$ involving
mean velocity $v_{m}$. Using these equations to express the two largest
coefficients $\alpha_{N-1,N}$ and $\beta_{N-1,N}$ as functions of
the lower order variables allows to formulate a reduced $N(\star)=N-2$
moment approximation. The projection of the Newtonian friction terms
$\partial_{\zeta\zeta}u$ and $\partial_{\zeta\zeta}v$ can then be
directly evaluated without partial integration. 

This approach also comes with disadvantages. The elimination of variables
strongly increases the nonlinearity of the equations making it difficult
to to write an explicit cascading form of a generic $N(\star)$-system.
Also, an explicit space dependence is introduce into the flux function,
since the constraint equations involve the possibly space dependent
slip-length $\lambda$ in general. Additionally, the direct projection
of the friction will in general not yield symmetric positive definite
matrices as in the weak case. However, an increased approximation
quality for cases with friction can be expected.

We evaluate conditions (\ref{eq:constraints}) for $N=3$ in one space
dimension and as before use $s=\alpha_{1}$, $\kappa=\alpha_{2}$,
and $m=\alpha_{3}$ for the coefficients. The explicit relations
for the higher order moments then read 
\begin{align}
\kappa=-\frac{5s+6u_{m}}{9(1+8\lambda/h)},\qquad & m=-\frac{4s(1+3\lambda/h)+3u_{m}}{9(1+8\lambda/h)}\label{eq:cubicConstraints}
\end{align}
which we will further simplify to the non-slip case by setting $\lambda=0$.
Using these no-slip relations in the cubic system (\ref{eq:cubic})/(\ref{eq:cubicdetails})
and droping the two highest equations, we arrive at the relatively
compact cubic($\star$) no-slip moment system
\begin{align}
\partial_{t}\left(\hspace{-2mm}\begin{array}{c}
h\\
hu_{m}\\
hs
\end{array}\hspace{-2mm}\right)+\partial_{x}\left(\hspace{-2mm}\begin{array}{c}
hu_{m}\\
\frac{116}{105}hu_{m}^{2}+\frac{80}{189}hs^{2}+\frac{4}{21}hu_{m}s+g\frac{h^{2}}{2}\\
\frac{4}{35}hu_{m}^{2}-\frac{20}{63}hs^{2}+\frac{12}{7}hu_{m}s
\end{array}\hspace{-2mm}\right)=\left(\hspace{-2mm}\begin{array}{c}
0\\
0\\
\frac{8u_{m}}{7}\partial_{x}(hs)-\frac{s}{7}\partial_{x}(hu_{m})
\end{array}\hspace{-2mm}\right)-P\label{eq:cubicStar}
\end{align}
using only the moment variable $s$ next to height $h$ and mean velocity
$u_{m}$. The direct evaluation of the friction terms based on the
constrained cubic polynomial yields the expression 
\begin{align}
P & =\frac{\nu}{h}\left(\begin{array}{c}
0\\
8u_{m}+\frac{20}{3}s\\
20u_{m}+\frac{80}{3}s
\end{array}\right)
\end{align}
for the production. 

\begin{figure}
\noindent \includegraphics[width=15cm]{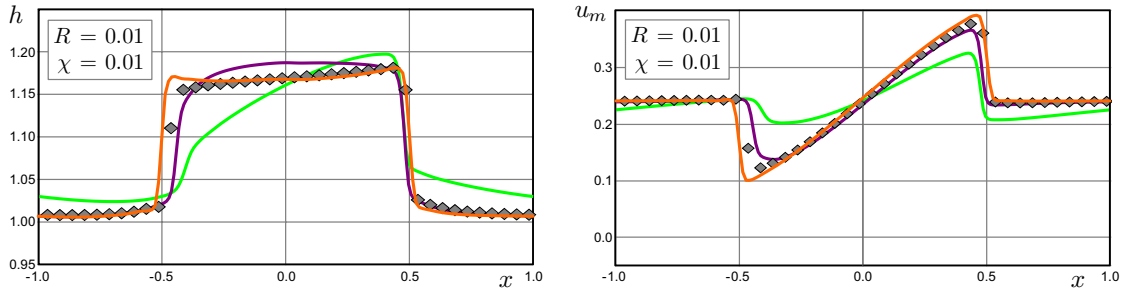}\caption{Moment approximations with strongly enforced no-slip boundary conditions
in the underlying vertical velocity profile (\textcolor{orange}{orange
lines}). Test case of \emph{viscous} unsteady solutions with linear
initial velocity profiles at $t_{\text{end}}=2$ and friction parameters
$R=\chi=0.01$, see Table \ref{table:setup}. Symbols represent the
result of the vertically resolved reference simulation, while the
other lines represent the \textcolor{green}{linear} (\textcolor{green}{green})
and \textcolor{violet}{cubic} (\textcolor{violet}{purple}) moment
system with weak boundary conditions (as in Fig.~\ref{fig:MomentsFriction})}
\label{fig:MomentsBCbuildin} 
\end{figure}

This cubic($\star$) no-slip system is based on a third degree polynomial
for the vertical velocity which will automatically satisfy free-force
conditions at the surface and vanishing velocity at the bottom. The
remaining flexibility of the profile is encoded into the mean velocity
and the linear moment $s$. In Fig.~\ref{fig:MomentsBCbuildin} we
show the result of the system for the test case of Table \ref{table:setup}
with friction coefficient $R=0.01$ and dimensionless slip length
$\chi=0.01$. Note that the cubic($\star$) system (\ref{eq:cubicStar})
formally assumes $\chi=0$, but is used here in an approximative way.
The figure shows good approximation of the reference data. In particular,
the quality of the result is clearly better than the original linear
system which uses the same number of equations and better than the
original cubic system with weak boundary conditions which uses the
same underlying polynomial degree. Note, that using the constraint
equations (\ref{eq:cubicConstraints}) with the precise slip length
$\chi=0.01$ would not show any visible improvement in the result.
Further approximation quality can be achieved by increasing the underlying
polynomial degree together with the contrainst (\ref{eq:constraints}),
which is left to future work.

\section{Conclusions}

In this article, we investigated mathematical models for shallow flow
systems. We conducted a scaling analysis of the fundamental balance
laws that reflects the process's shallow geometry and allows to identify
dominant terms. We formulated the resulting mathematical equations
as a vertically resolved reference system, which gave us direct access
to deriving a vertical model cascade for shallow flows based on moment
approximations. The model cascade was exemplified for purely kinematic,
nonviscous conditions as well as for Newtonian flow, in which a slip boundary 
condition at the bottom topography has been weakly enforced. Finally, we presented
numerical results in 1D for both the kinematic, and for the Newtonian
flow situation. The performance of the shallow flow moment cascade up to third
order has been investigated with respect to both the classical shallow
flow theory, and with respect to the fully vertically resolved reference
model. In a final example, we consider a shallow flow moment variant that 
strongly enforces the basal and free-surface boundary conditions. 
We found that many shallow flow regimes cannot be adequately
captured with the standard shallow flow theory. This is typically
true for a flow situation that is governed by a complex velocity profile,
e.g. induced by friction or by forced in-flow. By using
shallow flow moment systems, we can significantly improve the validity
of the mathematical model. Our test cases consistently showed a converging
behavior of the moment hierarchy. 

Similar to the classical shallow flow equations, the shallow moment
hierarchy makes use of the complexity reducing framework of depth-averaging.
Other than the standard shallow flow theory, however, it preserves
information on the vertical flow structure similar to a fully vertically
resolved mathematical model, yet being computationally less costly.
For physical flow regimes, in which the classical shallow flow model
results in a large model error and fails, it constitutes a powerful
alternative to improve on the predictive power of shallow flow simulations
in the future.

\appendix

\section{Dimensional Analysis\label{sec:derivation}}

We consider incompressible flow in three dimensions, for which mass
and momentum can be written as
\begin{align}
\nabla\cdot\mathbf{u} & =0\label{eq:massbal}\\
\partial_{t}\mathbf{u}+\nabla\cdot(\mathbf{u}\mathbf{u}) & =-\frac{1}{\rho}\nabla p+\frac{1}{\rho}\nabla\cdot\sigma+\mathbf{g}\label{eq:mombal}
\end{align}
The state variables velocity $\mathbf{u}=(u,v,w)^{T}$, pressure $p$
and deviatoric stress tensor $\sigma$ are functions of space $(x,y,z)$
and time $t$. The density $\rho$ is constant and $\mathbf{g}$ is
the vector of gravitational acceleration. We allow for a general situation,
in which the $z$-axis of the coordinate system is not necessarily
collinear with the vector of gravitational acceleration. This implies
$\mathbf{g}=[g_{x},g_{y},g_{z}]^{T}=g[e_{x},e_{y},e_{z}]^{T}$, in
which $[e_{x},e_{y},e_{z}]^{T}$ is a unit vector and $g$ the value
of gravitational acceleration. The often used shallow water coordinate
system is recovered by choosing $e_{x}=e_{y}=0$ and $\mathbf{g}=g[0,0,-1]^{T}$.
The flowing material is bounded by the basal topography $h_{b}(t,x,y)$
and the upper free surface $h_{s}(t,x,y)$. In the absence of mass
production at the ground, $h_{b}$ will be a function of $x$ and
$y$ alone and has no explicit time dependency. However, we formally
allow the basal topography to be time dependent, which allows to write
the kinematic boundary conditions for the material boundaries $h_{b}$
and $h_{s}$ in a symmetric way following \cite{pudasaini2007avalanche}
\begin{align}
\partial_{t}h_{s}+\left(u(t,x,y,h_{s}),v(t,x,y,h_{s})\right)^{T}\cdot\nabla h_{s} & =w(t,x,y,h_{s})\label{eq:kinematicbndrya}\\
\partial_{t}h_{b}+\left(u(t,x,y,h_{b}),v(t,x,y,h_{b})\right)^{T}\cdot\nabla h_{b} & =w(t,x,y,h_{b}).\label{eq:kinematicbndryb}
\end{align}
See also Fig.~\ref{fig:mapping} for a sketch of the situation
in one dimension.

\subsection{Scaling\label{sec:dimanalysis}}

We are interested in free-surface shallow flow, which is characterized
by a horizontal length scale $L$ that is much larger than the characteristic
vertical length scale $H$, hence $H/L=\epsilon<<1$. Consequently,
we scale
\begin{align}
x & =L\hat{x},\quad y=L\hat{y},\quad z=H\hat{z}.\label{eq:spatialscaling}
\end{align}
We furthermore assume a characteristic horizontal velocity scale given
by a generic velocity $U$. According to the shallowness the vertical
velocity will be much smaller, which suggests $\epsilon U$ as a characteristic
vertical velocity scale
\begin{align}
u & =U\hat{u},\quad v=U\hat{v},\quad w=\epsilon U\hat{w}.\label{eq:velscaling}
\end{align}
An appropriate time scale is then given by the ratio of spatial and
velocity scale
\begin{align}
t & =\frac{L}{U}\hat{t}.\label{eq:timescaling}
\end{align}
This scaling is motivated by our primary interest in velocity profiles
and $U$ can be specified later. An often used specific choice is
the free-fall time scale $T=\sqrt{L/g}$ which is based on $U=\sqrt{gL}$. 

Finally, we introduce characteristic stresses. Here, we assume that
the pressure scales with the hydrostatic pressure based upon the characteristic
height $H$ and stresses $\sigma$ with a characteristic stress $S$.
Furthermore, we assume that the basal shear stresses $\sigma_{xz}$
and $\sigma_{yz}$ are of larger order in the shallowness parameter
$\epsilon$ than lateral shear stresses $\sigma_{xy}$ and normal
shear stresses $\sigma_{kk},\,k\in\{x,y,z\}$, that is,
\begin{align}
p & =\rho gH\hat{p},\quad\sigma_{xz/yz}=S\hat{\sigma}_{xz/yz},\quad\sigma_{xx/xy/yy/zz}=\epsilon S\hat{\sigma}_{xx/xy/yy/zz}.\label{eq:stressscaling}
\end{align}
This is an appropriate assumption for many rheologies relevant to
our focus area, namely shallow flow in the geophysical context, as
these often scale linearly with the strain rate $\dot{\gamma}$. Upon
applying the shallow flow scaling (\ref{eq:spatialscaling}) and (\ref{eq:velscaling}),
we find for the the strain rate
\begin{align}
\dot{\gamma}=\frac{U}{H}\left(\begin{matrix}2\epsilon\partial_{\hat{x}}\hat{u} & \epsilon\left(\partial_{\hat{y}}\hat{u}+\partial_{\hat{x}}\hat{v}\right) & \partial_{\hat{z}}\hat{u}+\epsilon^{2}\partial_{\hat{x}}\hat{w}\\
\epsilon\left(\partial_{\hat{x}}\hat{v}+\partial_{\hat{y}}\hat{u}\right) & 2\epsilon\partial_{\hat{y}}\hat{v} & \partial_{\hat{z}}\hat{v}+\epsilon^{2}\partial_{\hat{y}}\hat{w}\\
\partial_{\hat{z}}\hat{u}+\epsilon^{2}\partial_{\hat{x}}\hat{w} & \partial_{\hat{z}}\hat{v}+\epsilon^{2}\partial_{\hat{y}}\hat{w} & 2\epsilon\partial_{\hat{z}}\hat{w}
\end{matrix}\right)
\end{align}
Neglecting all terms that scale with $\epsilon$ leads to the well-known
boundary layer approximation:
\begin{align}
\dot{\gamma}=\frac{U}{H}\left(\begin{matrix}0 & 0 & \partial_{\hat{z}}\hat{u}\\
0 & 0 & \partial_{\hat{z}}\hat{v}\\
\partial_{\hat{z}}\hat{u} & \partial_{\hat{z}}\hat{v} & 0
\end{matrix}\right)
\end{align}
For our purposes this is too restrictive, though. Instead, we will
keep terms of first order in $\epsilon$ but drop terms of second
order in $\epsilon^{2}$, which directly motivates the scaling introduced
in (\ref{eq:stressscaling}). The relevance of this assumption is
demonstrated by two examples: 
\begin{itemize}
\item \emph{Newtonian flow:} In the case of incompressible Newtonian shallow
flow, the deviatoric stress is given by
\begin{align}
\sigma=2\mu\dot{\gamma}=2\mu\frac{U}{H}\hat{\dot{\gamma}}.
\end{align}
With a characteristic stress defined as $S=\mu\frac{U}{H}$ we recover
the postulated scaling (\ref{eq:stressscaling}). 
\item \emph{Granular flow:} Similarly, the plastic deformation of dense
granular flow is governed by a constitute relation that scales linearly
with the strain-rate tensor. It is given by the so-called $\mu(I)$
rheology \cite{jop2006constitutive}, which is defined as
\begin{equation}
\sigma=\mu(I)\frac{p}{|\dot{\gamma}|}\dot{\gamma}
\end{equation}
Here, $\mu(I)$ is a generalized friction coefficient formulated in
terms of the inertial number $I$, the ratio of macroscopic deformation
time scale and grain inertial time scale. Furthermore, $|\dot{\gamma}|$
stands for the second invariant of the strain rate tensor $\dot{\gamma}$.
It has been shown that the $\mu(I)$-rheology is well-posed only for 
intermediate values of the inertial number \cite{BarkerSBG2015}, while a depth-averaged
formulation like \cite{gray2014depth} have much better properties.
Generally, this invariant is likewise affected by the shallow flow scaling (\ref{eq:spatialscaling})
and (\ref{eq:velscaling}), but has leading terms of zero order in
$\epsilon$:
\begin{equation}
|\dot{\gamma}|=\frac{U^{2}}{4H^{2}}\left((\partial_{\hat{z}}\hat{u})^{2}+(\partial_{\hat{z}}\hat{v})^{2}\right)+\mathcal{O}(\epsilon^{2})
\end{equation}
Hence, characteristic stress $S$ defined as $\mu(I)\frac{p}{|\dot{\gamma}|}$
is independent of $\epsilon$ and we again recover the scaling postulated
in (\ref{eq:stressscaling}). 
\end{itemize}

\subsection{Dimensionless System}

All in all, the componentwise mass and momentum equations written
in scaled dimensionless coordinates read
\begin{align}
\partial_{\hat{x}}\hat{u}+\partial_{\hat{y}}\hat{v}+\partial_{\hat{z}}\hat{w} & =0\\
F^{2}\epsilon\left(\partial_{\hat{t}}\hat{u}+\partial_{\hat{x}}\hat{u}^{2}+\partial_{\hat{y}}(\hat{u}\hat{v})+\partial_{\hat{z}}(\hat{u}\hat{w})\right) & =-\epsilon\partial_{\hat{x}}\hat{p}+\epsilon^{2}G\partial_{\hat{x}}\hat{\sigma}_{xx}+\epsilon G\partial_{\hat{y}}\hat{\sigma}_{xy}+G\partial_{\hat{z}}\hat{\sigma}_{xz}+e_{x}\\
F^{2}\epsilon\left(\partial_{\hat{t}}\hat{v}+\partial_{\hat{x}}(\hat{u}\hat{v})+\partial_{\hat{y}}(\hat{v}^{2})+\partial_{\hat{z}}(\hat{v}\hat{w})\right) & =-\epsilon\partial_{\hat{y}}\hat{p}+\epsilon G\partial_{\hat{x}}\hat{\sigma}_{xy}+\epsilon^{2}G\partial_{\hat{y}}\hat{\sigma}_{yy}+G\partial_{\hat{z}}\hat{\sigma}_{yz}+e_{y}\\
F^{2}\epsilon^{2}\left(\partial_{\hat{t}}\hat{w}+\partial_{\hat{x}}(\hat{u}\hat{w})+\partial_{\hat{y}}(\hat{v}\hat{w})+\partial_{\hat{z}}\hat{w}^{2}\right) & =-\partial_{\hat{z}}\hat{p}+\epsilon G\partial_{\hat{x}}\hat{\sigma}_{xz}+\epsilon G\partial_{\hat{y}}\hat{\sigma}_{yz}+\epsilon G\partial_{\hat{z}}\hat{\sigma}_{zz}+e_{z}
\end{align}
in which $F=U/\sqrt{gH}$ stands for the Froude number and $G=S/\rho gH$
for the ratio between characteristic stress and characteristic hydrostatic
pressure. Note that for a Newtonian constitutive relation ($S=\mu U/H$)
and a Froude Number close to one ($U^{2}\approx gH$), the parameter
$G$ reduces to the inverse of the well-known Reynolds Number, namely
$G=\nu/(HU)$, with $\nu$ standing for the kinematic viscosity $\mu/\rho$.

\noindent In regimes in which the stress is dominated by hydrostatic
pressure we have $G<1$, and hence contributions of the order $\epsilon^{2}$
and $\epsilon G$ will be insignificant. The system can then be reduced
to
\begin{align}
\partial_{\hat{x}}\hat{u}+\partial_{\hat{y}}\hat{v}+\partial_{\hat{z}}\hat{w} & =0\label{eq:massbalcompdim}\\
F^{2}\epsilon\left(\partial_{\hat{t}}\hat{u}+\partial_{\hat{x}}\hat{u}^{2}+\partial_{\hat{y}}(\hat{u}\hat{v})+\partial_{\hat{z}}(\hat{u}\hat{w})\right) & =-\epsilon\partial_{\hat{x}}\hat{p}+G\partial_{\hat{z}}\hat{\sigma}_{xz}+e_{x}\label{eq:momxbalcompdim}\\
F^{2}\epsilon\left(\partial_{\hat{t}}\hat{v}+\partial_{\hat{x}}(\hat{u}\hat{v})+\partial_{\hat{y}}(\hat{v}^{2})+\partial_{\hat{z}}(\hat{v}\hat{w})\right) & =-\epsilon\partial_{\hat{y}}\hat{p}+G\partial_{\hat{z}}\hat{\sigma}_{yz}+e_{y}\label{eq:momybalcompdim}\\
0 & =-\partial_{\hat{z}}\hat{p}+e_{z}.\label{eq:momzbalcompdim}
\end{align}

\noindent From the reduced vertical momentum balance (\ref{eq:momzbalcompdim}),
we can directly solve for the pressure profile
\begin{align}
\hat{p}(\hat{t},\hat{x},\hat{y},\hat{z})=\left(\hat{h}_{s}(\hat{t},\hat{x},\hat{y})-\hat{z}\right)e_{z}\label{eq:pressure}
\end{align}
which satisfies a stress-free boundary condition at the free surface
$\hat{p}(\hat{t},\hat{x},\hat{y},\hat{h}_{s}(\hat{t},\hat{x},\hat{y}))=0$.
This system forms the starting point of Sec.\,\ref{sec:landau} in
the main text.

\section{Higher Order Moment Equations\label{sec:Higher-Order-Moment}}

For completeness we present the details of deriving the equation (\ref{eq:highermoments-x})
for the higher order expansion coefficients $\alpha_{i}$. Defining
\begin{align}
A_{ijk}=(2i+1)\int_{0}^{1}\phi_{i}\,\phi_{j}\,\phi_{k}\,d\zeta\label{eq:Aijk}
\end{align}
we can write 
\begin{align}
\int_{0}^{1}\phi_{i}u^{2}d\zeta & =\tfrac{1}{2i+1}(2u_{m}\alpha_{i}+\sum_{j,k=1}^{N}A_{ijk}\alpha_{j}\alpha_{k})
\end{align}
and 
\begin{align}
\int_{0}^{1}\phi_{i}uvd\zeta & =\tfrac{1}{2i+1}(u_{m}\beta_{i}+v_{m}\alpha_{i}+\sum_{j,k=1}^{N}A_{ijk}\alpha_{j}\beta_{k})
\end{align}
for the moments of the convective terms in (\ref{eq:hydbalmomx}).
For the vertical coupling term we define 
\begin{align}
B_{ijk}=(2i+1)\int_{0}^{1}\phi_{i}'\,\left(\int_{0}^{\zeta}\phi_{j}\,d\hat{\zeta}\right)\,\phi_{k}\,d\zeta\label{eq:Aij(k)}
\end{align}
such that we can write 
\begin{align}
 & \int_{0}^{1}\phi_{i}\,\partial_{\zeta}\left(hu\omega\left(h,u\right)\right)d\zeta\\
 & =-\int_{0}^{1}\phi_{i}\,\partial_{\zeta}\left((u_{m}+u_{d})\left(\partial_{x}\left(h\int_{0}^{\zeta}u_{d}d\hat{\zeta}\right)+\partial_{y}\left(h\int_{0}^{\zeta}v_{d}d\hat{\zeta}\right)\right)\right)d\zeta\nonumber \\
 & =-u_{m}\,\partial_{x}\left(h\int_{0}^{1}\phi_{i}u_{d}d\zeta\right)-u_{m}\partial_{y}\left(h\int_{0}^{1}\phi_{i}v_{d}d\zeta\right)\nonumber \\
 & \quad+\int_{0}^{1}(\partial_{\zeta}\phi_{i})\,u_{d}\partial_{x}\left(h\int_{0}^{\zeta}u_{d}d\hat{\zeta}\right)d\zeta+\int_{0}^{1}(\partial_{\zeta}\phi_{i})u_{d}\partial_{y}\left(h\int_{0}^{\zeta}v_{d}d\hat{\zeta}\right)d\zeta\nonumber \\
 & =-\tfrac{1}{2i+1}u_{m}\left(\partial_{x}\left(h\alpha_{i}\right)+\partial_{y}\left(h\beta_{i}\right)\right)+\tfrac{1}{2i+1}\sum_{j=1}^{N}\sum_{k=1}^{N}B_{ijk}\alpha_{k}\left(\partial_{x}\left(h\alpha_{j}\right)+\partial_{y}\left(h\beta_{j}\right)\right)\nonumber \\
 & =-\tfrac{1}{2i+1}u_{m}D_{i}+\tfrac{1}{2i+1}\sum_{j=1}^{N}\sum_{k=1}^{N}B_{ijk}\alpha_{k}D_{i}\nonumber 
\end{align}
where we used the definition (\ref{eq:omega}) for $\omega$ and $D_{i}$
for the divergence of $(\alpha_{i},\beta_{i})$ as in (\ref{eq:divD}).
Finally, we use 
\begin{align}
C_{ij}=\int_{0}^{1}\phi_{i}'\,\phi_{j}'\,d\zeta\label{eq:Aij}
\end{align}
for the friction term and obtain 
\begin{align}
\frac{\nu}{h}\int_{0}^{1}\phi_{i}\,\partial_{\zeta\zeta}u\,d\zeta & =\frac{\nu}{h}\left.\left(\phi_{i}\,\partial_{\zeta}u\right)\right|_{\zeta=0}^{\zeta=1}-\frac{\nu}{h}\int_{0}^{1}\phi_{i}'\,\partial_{\zeta}u\thinspace d\zeta\\
 & =-\frac{\nu}{\lambda}(u_{m}+\sum_{j=1}^{N}\alpha_{j})-\frac{\nu}{h}\sum_{j=1}^{N}C_{ij}\alpha_{j}.\nonumber 
\end{align}
The $y$-component (\ref{eq:highermoments-y}) is derived analogously
using the same coefficients $A_{ijk}$, $B_{ijk}$, and $C_{ij}$.

\section{Characteristic polynomial of the cubic moment system}
\label{charpolthird}

The eigenvalues of the flux Jacobian of the cubic moment system are of the form 
$a=u_{m}+c\sqrt{gh}$ where $c$ is the any root of the polynomial
\begin{eqnarray*}
\begin{aligned}
&  \qquad \qquad p(c) = c^5 + \gamma_4 c^4 + \gamma_3 c^3 + \gamma_2 c^2 + \gamma_1 c^1 + \gamma_0 \\[0.2cm] 
\end{aligned}
\end{eqnarray*}
with coeficients
\begin{eqnarray*}
\begin{aligned}
\gamma_4 &= -\frac{37 \kappa}{21} \\
\gamma_3 &= \frac{32 \kappa^2}{105}-\frac{4 m^2}{3}-\frac{4 m s}{3}-\frac{10s^2}{7}-1 \\
\gamma_2 &= \frac{24 \kappa^3}{35}+\frac{344 \kappa m^2}{735}-\frac{404 \kappa m s}{245}+
\frac{2 \kappa s^2}{5}+\frac{37 \kappa}{21} \\
	\gamma_1 &= \frac{5 \kappa^4}{21}-\frac{142 \kappa^2 m^2}{735}+\frac{192}{245} \kappa^2
   m s+\frac{8 \kappa^2 s^2}{35}\\
	& \quad  -\frac{19 \kappa^2}{21}+\frac{m^4}{3}-\frac{4 m^2 s^2}{7}+\frac{19
   m^2}{21}+\frac{4 m s^3}{21}+\frac{4 m s}{3}+\frac{3 s^4}{7}+\frac{3
   s^2}{7} \\
	\gamma_0 &= \frac{\kappa^5}{105}+\frac{22 \kappa^3 m^2}{735}+\frac{104}{735} \kappa^3 m s-\frac{8 \kappa^3
   s^2}{35}+\frac{\kappa^3}{7}-\frac{29 \kappa m^4}{735}-\frac{176}{735} \kappa m^3 s\\
	& \quad  -\frac{104}{735} \kappa m^2
   s^2-\frac{25 \kappa m^2}{147}+\frac{68}{245} \kappa m s^3-\frac{20 \kappa m s}{49}+\frac{23 \kappa s^4}{105}-\frac{5 \kappa
   s^2}{21}
\end{aligned}
\end{eqnarray*}
Note, that the variables $s, \kappa, m$ all have been scaled by $\sqrt{gh}$ for better readibility.

\bibliographystyle{siam}

\end{document}